\documentclass[twocolumn]{aastex62}
\usepackage{natbib}
\usepackage{amsfonts}
\usepackage{amsmath}
\usepackage{graphicx}
\usepackage{xspace}
\usepackage{mathrsfs}
\usepackage{bm}
\usepackage[normalem]{ulem}

\hypersetup{linkcolor=cyan,citecolor=cyan,filecolor=cyan,urlcolor=cyan}

\newcommand{\code}[1]{\texttt{#1}}

\newcommand\grad\nabla
\newcommand\grada{\nabla_{\rm ad}}
\renewcommand\bv{Brunt-V\"ais\"al\"a\xspace}
\newcommand\scvhi{SCvH-i\xspace}
\newcommand\emcee{\code{emcee}\xspace}
\newcommand\gyre{GYRE\xspace}

\newcommand\ompat{\Omega_{\rm pat}}
\newcommand\siglmn{\sigma_{\l mn}} 
\newcommand\siglmnc{\siglmn^{\rm corot}} 

\newcommand\tone{T_1}

\newcommand\gmsat{GM}
\newcommand\rsat{R}
\newcommand{\mc}{\ensuremath M_{\rm c}}
\newcommand\ptrans{P_{12}}
\newcommand\mrot{m_{\Omega}}

\newcommand\xir{\xi_r}
\newcommand\xih{\xi_h}
\newcommand\rhop{\rho^\prime}
\renewcommand\phi{\varphi}

\renewcommand\l\ell

\newcommand\omegasat{\Omega_{\rm S}}
\newcommand\psat{P_{\rm S}}
\newcommand\tildesigma{\tilde\sigma_{\l mn}}

\newcommand{\cassini}{\emph{Cassini}\xspace}
\newcommand{\voyager}{\emph{Voyager}\xspace}

\renewcommand{\d}[1]{\ensuremath{\operatorname{d}\!{#1}}} 

\newcommand\hnone{\cite{2013AJ....146...12H}}
\newcommand\hntwo{\cite{2014MNRAS.444.1369H}}
\newcommand\french{\cite{2016Icar..279...62F}}

\renewcommand\S{Section~}

\accepted{to ApJ} 

\begin{document}
\shortauthors{Mankovich et al.}
\shorttitle{Saturn's Seismological Rotation}
\title{\cassini Ring Seismology as a Probe of Saturn's Interior I: Rigid Rotation}

%

\correspondingauthor{Chris Mankovich}
\email{cmankovich@ucsc.edu}

\newcommand\ucsc{Department of Astronomy and Astrophysics, University of California Santa Cruz}
\newcommand\arc{NASA Ames Research Center}

\author{Christopher Mankovich}
\affiliation\ucsc

\author{Mark S. Marley}
\affiliation\arc

\author{Jonathan J. Fortney}
\affiliation\ucsc

\author{Naor Movshovitz}
\affiliation\ucsc

\begin{abstract}
    Seismology of the gas giants holds the potential to resolve long-standing questions about their internal structure and rotation state. We construct a family of Saturn interior models constrained by the gravity field and compute their adiabatic mode eigenfrequencies and corresponding Lindblad and vertical resonances in Saturn's C ring, where more than twenty waves with pattern speeds faster than the ring mean motion have been detected and characterized using high-resolution \cassini Visual and Infrared Mapping Spectrometer (VIMS) stellar occultation data. We present identifications of the fundamental modes of Saturn that appear to be the origin of these observed ring waves, and use their observed pattern speeds and azimuthal wavenumbers to estimate the bulk rotation period of Saturn's interior to be $10{\rm h}\, 33{\rm m}\, 38{\rm s}^{+1{\rm m}\, 52{\rm s}}_{-1{\rm m}\, 19{\rm s}}$ (median and 5\%/95\% quantiles), significantly faster than \voyager and \cassini measurements of periods in Saturn's kilometric radiation, the traditional proxy for Saturn's bulk rotation period. The global fit does not exhibit any clear systematics indicating strong differential rotation in Saturn's outer envelope.
\end{abstract}

\keywords{planets and satellites: individual (Saturn) -- planets and satellites: interiors -- planets and satellites: rings}

\section{Introduction}\label{s.intro}
The prototypical gas giants Jupiter and Saturn offer an opportunity to study the processes at work during planet formation and the chemical inventory of the protosolar disk, and also constitute astrophysical laboratories for warm dense matter. Inferences about these planets' composition and structure rely on interior models that are chiefly constrained by the their observed masses, radii and shapes, surface abundances, and gravity fields \citep{1982AREPS..10..257S,2016arXiv160906324F}. While the latter have been measured to unprecedented precision by \emph{Juno} at Jupiter and the \cassini Grand Finale at Saturn, in the interest of long term progress there is a need to identify independent observational means of studying the interiors, and seismology using the planets' free oscillations appears to be the most promising such avenue.

While preliminary detections of Jupiter's oscillations have been made from the ground by \cite{2011A&A...531A.104G}, the resulting power spectrum lacked the frequency resolution necessary to identify specific normal modes responsible for the observed power, a necessary step before the frequencies can be used to probe the interior in detail.
Saturn, on the other hand, provides a unique opportunity for seismic sounding of a Jovian interior owing to its highly ordered ring system, wherein gravity perturbations from Saturn's free oscillations can resonate with ring orbits. Saturn ring seismology is the focus of this work.

\subsection{Background}
\label{s.storytime}
The concept of ring seismology was first developed in the 1980s. \citet{stevenson1982} suggested that Saturnian inertial oscillation modes, for which the Coriolis force is the restoring force, could produce regular density perturbations within the planet that might resonate with ring particle orbits and open gaps or launch waves, but he did not calculate specific mode frequencies. Later in the decade in a series of abstracts, a thesis, and papers Marley, Hubbard and Porco further developed this idea. \citet{1987BAAS...19..889M}, relying on Saturn oscillation frequencies computed by \citet{1981SvA....25..724V}, suggested that acoustic mode oscillations, which differ from inertial modes in that their restoring force is ultimately pressure, could resonate with ring particle orbits in the C ring. They recognized that mode amplitudes of a few meters would be sufficient to perturb the rings. \citet{1988BAAS...20Q.870M} focused on low angular degree $\l$ $f$-modes which have no radial nodes in displacement from surface to the center of the planet (unlike $p$-modes) as the modes which had the potential to provide the most information about the deep interior of a giant planet. \citet{1989BAAS...21..928M} compared the predicted locations resonance locations of such modes with newly discovered wave features in the C ring found in radio occupation data by \citet{1989PhDT.........2R}. They suggested that the Maxwell gap and three wave features found by Rosen which had azimuthal wave numbers and propagation directions consistent with such resonances were in fact produced by Saturnian $f$-modes with $\l\le4$. As we will summarize below, we now know that these specific $f$-mode--ring feature associations were correct, although the story for the $\l=2$ and $\l=3$ waves is complicated by $g$-mode mixing \citep{2014Icar..231...34F,2014Icar..242..283F}.

These ideas were ultimately presented in detail in \citet{1990PhDT.........3M,1991Icar...94..420M} and \citet{1993Icar..106..508M}. Marley computed the sensitivity of Saturn oscillation frequencies to various uncertainties in Saturn interior models, including core size and regions with composition gradients, and discussed the sensitivity of ring resonance locations to these uncertainties. As we will show below, the overall pattern of resonance locations within the rings first presented in \citet{1990PhDT.........3M} agrees well with subsequent discoveries. While Marley recognized the impact of regions with non-zero \bv frequency $N$ on $f$-mode frequencies and the possibility of $g$-modes (for which the restoring force is buoyancy), he did not consider mode mixing between $f$- and $g$-modes. \citet{1993Icar..106..508M} presented the theory of resonances between planetary oscillation modes and rings in detail and derived expressions for the torque applied to the rings at horizontal (Lindblad) and vertical resonances and compared these torques to those of satellites. They also suggested several more specific ring feature-oscillation mode associations, many of which have subsequently turned out to be correct. Marley and Porco concluded by noting that because the azimuthal wave numbers of the Rosen wave features were uncertain, only additional observations by the planned future Saturn mission \cassini could ultimately test the hypothesized  oscillation mode--ring feature connection. Consequently there was an essentially two-decade pause in ring seismology research until those results became available.

Optical depth scans of the C ring from \cassini radio occultations and Ultraviolet Imaging Spectrograph stellar occultations presented by \citet{2009sfch.book..375C} and \citet{2011Icar..216..292B} confirmed all the unexplained waves reported by \citet{1991Icar...93...25R} and identifying many more. \citet{2013AJ....146...12H} followed up with VIMS stellar occultations, combining scans taken by \cassini at different orbital phases to determine wave pattern speeds and azimuthal wavenumbers $m$ at outer Lindblad resonances, making seismology of Saturn using ring waves possible for the first time. As alluded to above, the  detection of multiple close waves with $m=2$ and $m=3$ waves deviated from the expectation for the spectrum of pure $f$-modes. In light of this result, \citet{2014Icar..231...34F} investigated the possibility of shear modes in a solid core, finding that rotation could mix these core shear modes with the $f$-modes and in principle explain the observed fine splitting, although they noted that some fine tuning of the model was required. The most compelling model for the fine splitting to date was presented by \citet{2014Icar..242..283F}, who showed that a strong stable stratification outside Saturn's core would admit $g$-modes that could rotationally mix with the $f$-modes and rather robustly explain the number of strong split $m=2$ and $m=3$ waves at Lindblad resonances, and roughly explain the magnitude of their frequency separations.

Subsequent obervational results from the VIMS data came from \citet{2014MNRAS.444.1369H}, who detected a number of additional waves including an $m=10$ wave apparently corresponding to Saturn's $\l=m=10$ $f$-mode. \citet{2016Icar..279...62F} characterized the wave in the ringlet within the Maxwell gap \citep{2005Sci...307.1226P} and argued it to be driven by Saturn's $\l=m=2$ $f$-mode, supporting the prediction by \citet{1989BAAS...21..928M}.
The remainder of C ring wave detections that form the observational basis for our work are the density waves reported by \cite{2018arXiv181104796H} and the density and bending waves reported by \cite{rgf2018}.

\subsection{This work}
Here we seek to systematically understand the ring wave patterns associated with Saturn's normal modes. In particular, we aim to identify the modes responsible for each wave, make predictions for the locations of other Saturnian resonances in the rings, and ultimately assess what information these modes carry about Saturn's interior.
We describe the construction of Saturn interior models in Section \ref{s.models}.
\S\ref{s.eigenstuff} summarizes our method for solving for mode eigenfrequencies and eigenfunctions, as well as our accounting for Saturn's rapid rotation. In \S\ref{s.resonances} we recapitulate the conditions for Lindblad and vertical resonances with ring orbits and describe which $f$-modes can excite waves at each.
\S\ref{s.results} presents the main results, namely $f$-mode identifications and a systematic comparison of predicted $f$-mode frequencies to the pattern speeds of observed waves and its implications for Saturn's interior, principally its rotation.
The separate question of mode amplitudes and detectability of ring waves is addressed in \S\ref{s.torques}, which also lists the strongest predicted waves yet to be detected.
Discussion follows in \S\ref{s.discussion} and we summarize our conclusions with \S\ref{s.conclusion}.

\section{Interior Models}\label{s.models}
Our hydrostatic planet interior models are computed using a code based on that of \cite{2016ApJ...831...64T} with a few important generalizations.
To model arbitrary mixtures of hydrogen and helium, we implement the equation of state of \cite{1995ApJS...99..713S} (the version interpolated over the plasma phase transition, henceforth ``\scvhi''). Heavier elements are included using the ab initio water EOS of \cite{2009PhRvB..79e4107F}, extending the coverage to $T<10^3$ K using the analytical model of \cite{aneos} for water. The density $\rho(Y,Z)$ is obtained assuming linear mixing of the three components following
\begin{equation}
    \rho^{-1}(Y,Z)=\frac{Z}{\rho_Z}+\frac{1-Z}{\rho_{\rm HHe}(Y)},
\end{equation}
where in turn
\begin{equation}
    \rho_{\rm HHe}^{-1}(Y)=\frac{Y}{\rho_{\rm He}}+\frac{1-Y}{\rho_{\rm H}}.
\end{equation}
Here $Y$ and $Z$ are the mass fractions of helium and heavier elements, respectively, and the densities $\rho_{\rm H}$, $\rho_{\rm He}$, and $\rho_{\rm Z}$ are tabulated as functions of pressure $P$ and temperature $T$ in the aforementioned equations of state.

The outer boundary condition for our interior models is simply a fixed temperature at $P=1\ {\rm bar}$, namely $\tone=140\ {\rm K}$, close to the value derived by \cite{1985AJ.....90.1136L} from \voyager radio occultations and mirroring that used in previous Saturn interior modeling efforts (e.g., \citealt{2013Icar..225..548N}). The envelope is assumed to be everywhere efficiently convective so that the deeper temperature profile is obtained by integrating the adiabatic temperature gradient:
\begin{equation}
    \label{eq.integrate_grada}
    T(m_r\geq\mc)=T_1 + \int_M^{m_r}\grada(P,\,T,\,Y)\,T\,\d{\ln P},
\end{equation}
with the core itself assumed isothermal at $T(\mc)$.
Here $m_r$ denotes the mass coordinate and the adiabatic temperature gradient $\grada\equiv\left(\frac{\partial\ln T}{\partial \ln P}\right)_{\rm ad}$ is assumed to be that of the hydrogen-helium mixture alone
\footnote{
This simplification is necessary because the water tables of \cite{2009PhRvB..79e4107F} do not provide an entropy column. While these tables have been extended with entropies calculated from separate thermodynamic integrations (N.~Nettelmann, private communication), the entropies are accurate only up to an additive offset and so cannot be used to write the total entropy of even an ideal H-He-Z mixture. Within the core where $Z=1$, the entropy is straightforward to calculate and there we use these extended tables to calculate the sound speed in pure water. See \cite{2008A&A...482..315B} for a discussion of the significance of heavy elements in setting $\grada$ in the envelope.
}.

Following common choices for models of Saturn's interior \citep[e.g.,][]{2013Icar..225..548N}, the distribution of constituent species with depth follows a three-layer piecewise homogeneous structure: heavy elements are partitioned into a core devoid of hydrogen and helium ($Z=1$) and a two-layer envelope with outer (inner) heavy element mass fraction $Z_1$ ($Z_2$). The helium content is likewise partitioned with outer (inner) helium mass fraction $Y_1$ ($Y_2$) subject to the constraint that the mean helium mass fraction of the envelope match the protosolar nebula abundance $Y=0.275$. The $Z$ and $Y$ transitions are located at a common pressure level $\ptrans$, a free parameter conceptually corresponding to the molecular-metallic transition of hydrogen, although in \scvhi itself this is explicitly a smooth transition. We only consider $Z_2>Z_1$ and $Y_2>Y_1$ to avoid density inversions and to reflect the natural configuration of a differentiated planet.

The particular choice of this three-layer interior structure model is motivated by the desire for a minimally complicated model that simultaneously (a) satisfies the adopted physically-motivated EOS, (b) includes enough freedom to fit Saturn's low-order gravity field $J_2$ and $J_4$, and (c) does not introduce significant convectively stable regions in the envelope, such as those that might arise in cases where composition varies continuously. Requirement (c) precludes a viable class of configurations for Saturn's interior (e.g., \citealt{2013NatGe...6..347L}, \citealt{2014Icar..242..283F}, \citealt{2016ApJ...829..118V}), but it significantly simplifies the formalism and interpretation because in this case the normal modes in the relevant frequency range are limited to the fundamental and acoustic overtone modes.
While the isothermal cores of our models are stably stratified and so do admit $g$-modes, the stratification is such that the maximum \bv frequency attained there is only $N\approx\sigma_0$, where $\sigma_0=(\gmsat/\rsat)^{1/2}$ is Saturn's dynamical frequency. Since $g$-modes have frequencies at most $N$, and $f$-mode frequencies follow $\sigma\approx\l^{1/2}\sigma_0$ \citep{1980LNP...125....1G}, $g$-modes in such a core will not undergo avoided crossings with the $\l\geq2$ $f$-modes.
As will be discussed in \S\ref{s.results} below, a spectrum of purely acoustic modes is sufficient to explain the majority of the spiral density and bending waves identified in the C ring that appear to be Saturnian in origin.

\subsection{Gravity field}
\label{s.gravity}
We generate rigidly rotating, oblate interior models by solving for the shape and mass distribution throughout the interior using the theory of figures formalism \citep{1978ppi..book.....Z}. The theory of figures expresses the total potential, including gravitational and centrifugal terms, as a series expansion in the small parameter $\mrot=\Omega^2R^3/GM$ where $\Omega$ is the uniform rotation rate, $R$ is the planet's volumetric mean radius, and $GM$ is the planet's total gravitational mass. Retaining terms of $\mathcal O(\mrot^n)$ provides a system of $n$ algebraic equations that describe the shape and total potential as integral functions of the two-dimensional mass distribution, while the mass distribution is in turn related to the potential by the condition of hydrostatic balance. A self-consistent solution for the shape and mass distribution in the oblate model is obtained iteratively, yielding the corresponding gravitational harmonics $J_{2n}$ in the process.
To this end we use the shape coefficients given through $\mathcal{O}(m^4)$ by \cite{2017A&A...606A.139N} and implement a similar algorithm. For our Saturn models we adopt $R=58,232\ {\rm km}$ \citep{2007CeMDA..98..155S} and
$\gmsat=37,931,207.7\ {\rm cm}^{3}\ {\rm s}^{-2}$ \citep{2006AJ....132.2520J}.

For a given combination of the parameters $Z_1$, $Z_2$, $Y_1$, $\ptrans$, and $\mrot$, an initially spherical model is relaxed to its rotating hydrostatic equilibrium configuration. The mean radii of level surfaces are adjusted during iterations such that the equatorial radius $a$ of the outermost level surface for a converged model matches $a=60,260$ km following \cite{2007CeMDA..98..155S}. As the mean radii are adjusted and the densities are recalculated from the EOS, the total mass of the model necessarily changes; therefore the core mass $\mc$ is simultenously adjusted over the course of iterations such that the converged model matches Saturn's total mass. These models include 4096 zones, the algorithm adding zones late in iterations if necessary to speed convergence to the correct total mass.

\begin{figure}[t]
    \begin{center}
        \includegraphics[width=1.0\columnwidth]{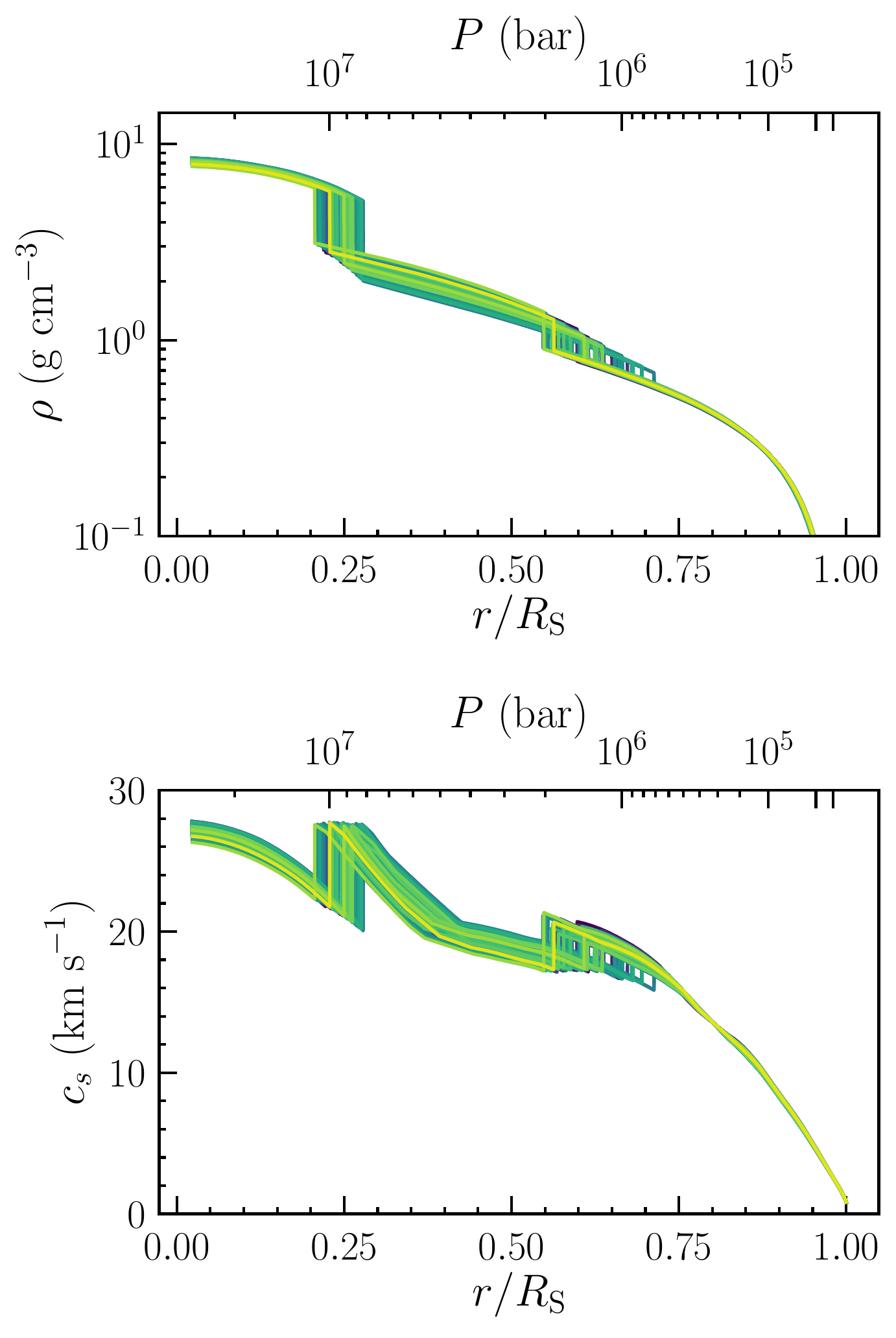}
        \caption{
        Saturn interior models with two-layer envelopes of varying $Y$ and $Z$ distributions, surrounding pure-$Z$ cores. Models are sampled based on $J_2$ and $J_4$ from \cite{iess2018}. Mass density (top panel) and sound speed (bottom panel) are shown as functions of the mean radii of level surfaces (bottom horizontal axes) and pressure coordinate (top horizontal axes).
        }
        \label{fig.profiles_compare_interiors}
    \end{center}
\end{figure}

The values for the gravity used for generating interior models are those of \cite{2006AJ....132.2520J}, appropriately normalized to our slightly smaller adopted reference equatorial radius according to $J_{2n}^\prime=(a/a^\prime)J_{2n}$. Although dramatically more precise harmonics obtained from the \cassini Grand Finale orbits will soon be published, the values of $J_2$ and $J_4$ from \cite{2006AJ....132.2520J} are already precise to a level beyond that which can be used to put meaningful constraints on the deep interior using our fourth-order theory of figures, where in practice solutions are only obtained with numerical precision at the level of $|\delta J_2/J_2|\approx|\delta J_4/J_4|\lesssim10^{-4}$.

For the purpose of fitting the gravity field, we create models using $\mrot=0.13963$ corresponding to the 10h 39m 24s $(10.657{\rm h})$ rotation period measured from \voyager kilometric radiation and magnetic field data by \cite{1981GeoRL...8..253D}. We sample interior models from a bivariate normal likelihood distribution in $J_2$ and $J_4$ using \emcee \citep{2013PSP..125..306F} assuming a diagonal covariance for these gravity harmonics. Because the numerical precision to which our theory of figures can calculate $J_2$ exceeds its observational uncertainty, the former is used in our likelihood function.
We take uniform priors on $Z_1$ and $Z_2$ subject to the constraint that $0<Z_1<Z_2<1$, a uniform prior on $0<Y_1<0.275$, and a uniform prior over $0.5\ {\rm Mbar}<\ptrans<2\ {\rm Mbar}$. The mass distributions and sound speeds for models in this sample are illustrated in Figure~\ref{fig.profiles_compare_interiors}.

\section{Mode eigenfrequencies and eigenfunctions}\label{s.eigenstuff}
Our approach is to perform the pulsation calculation for spherical models corresponding to the converged theory of figures models, with the various material parameters defined on the mean radii $r$ of level surfaces. The influence of Saturn's rapid rotation is accounted for after the fact using a perturbation theory that expresses the full solutions in the presence of Coriolis and centrifugal forces and oblateness in terms of linear superpositions of the solutions obtained in the non-rotating case.

For spherical models, we solve the fourth-order system of equations governing linear, adiabatic oscillations \citep{1989nos..book.....U} using the open source \gyre stellar oscillation code suite \citep{2013MNRAS.435.3406T}. The four assumed boundary conditions correspond to the enforcement of regularity of the eigenfunctions at $r=0$ and the vanishing of the Lagrangian pressure perturbation at the planet's surface $r=R$ \citep[Section 18.1]{1989nos..book.....U}.
The three-layer nature of the interior models considered in this work involve two locations at which the density and sound speed are discontinuous as a result of discontinuous composition changes (see Figure~\ref{fig.profiles_compare_interiors}). Additional conditions are applied at the locations of these discontinuities; these amount to jump conditions enforcing the conservation of mass and momentum across these boundaries.

As will be discussed in \S\ref{s.results}, comparison with the full set of observed waves in the C ring requires $f$-modes with angular degree in the range $\l=2-14$, and we tabulate results through for the $f$-modes through $\l=15$.

In what follows, we adopt the convention that $m>0$ corresponds to prograde modes---those that propagate in the same sense as Saturn's rotation---so that the time-dependent Eulerian perturbation to, e.g., the mass density corresponding to the $\l mn$ normal mode in the planet is written as
\begin{equation}
    \label{eq.eigenfunctions}
    \rho_{\l mn}^\prime(r,\theta,\phi,t)=\rho_{\l mn}^\prime(r)Y_\l^m(\theta,\phi)e^{-i\siglmn t},
\end{equation}
where $\siglmn$ is the mode frequency in the frame rotating with the planet, and $r$, $\theta$, and $\phi$ denote radius, colatitude, and azimuth respectively. Analogous relations hold with the pressure $P$ or gravitational potential $\Phi$ in place of density. The $Y_\l^m(\theta,\phi)$ are the spherical harmonics, here defined in terms of the associated Legendre polynomials $P_\l^m$ as
\begin{equation}
    \label{eq.spherical_harmonics}
    \begin{split}
    Y_\l^m(\theta,\phi) = &
    (-1)^{\frac{m+|m|}{2}}
    \left[
    \left(\frac{2\l+1}{4\pi}\right)
    \left(\frac{(\l-|m|)!}{(\l+|m|)!}\right)
    \right]^{1/2} \\
    &\quad\times
    P_\l^m(\cos\theta)e^{im\phi}.
    \end{split}
\end{equation}
The solution for the displacement itself has both radial and horizontal components, with the total displacement vector given by
\begin{equation}
    \label{eq.displacement_eigenfunctions}
    \begin{split}
    \bm\xi(r, \theta, \phi, t) = &\left[
    \xir(r)\,\hat{r}
    +\xih(r)
    \left(
    \hat\theta\frac{\partial}{\partial\theta}+\hat\phi\frac{1}{\sin\theta}\frac{\partial}{\partial\phi}
    \right)
    \right] \\
    &\quad \times Y_\l^m(\theta,\phi)e^{-i\sigma_{\ell mn}t}.
    \end{split}
\end{equation}

\subsection{Rotation}
\label{s.rotation_corrections}
In reality, Saturn's eigenfrequencies are significantly modified by the action of Saturn's rapid rotation because of Coriolis and centrifugal forces and the ellipticity of level surfaces. We account for these following the perturbation theory given by \cite{1981SvA....25..627V} (see also \citealt{1981ApJ...244..299S}) and later generalized by \cite{1981SvA....25..724V} to treat differential rotation, using the eigenfunctions obtained in the non-rotating case as basis functions for expressing the full solutions. In this work we calculate corrected eigenfrequencies for a range of rotation rates, treating Saturn as a rigidly rotating body.

Denoting by $\tildesigma$ the eigenfrequency obtained for the $\l mn$ mode in the non-rotating case, we write the corrected eigenfrequency as an expansion to second order in the small parameter
\begin{equation}
    \label{eq.define_lambda}
    \lambda\equiv\frac{\omegasat}{\tildesigma}
\end{equation}
so that the corrected frequency as seen in inertial space is given by
\begin{equation}
    \label{eq.correction_expansion}
    \siglmn=\tildesigma\left[1+\sigma_{\l mn,1}\lambda + \sigma_{\l m n,2}\lambda^2 + \mathcal O(\lambda^3)\right].
\end{equation}
For Saturn's $f$-modes, $\lambda\approx0.3$ for $\l=2$ and decreases to $\lambda\approx0.1$ by $\l=15$.
The dimensionless factor $\sigma_{\l mn,1}$ includes the effects of the Coriolis force and the Doppler shift out of the planet's rotating reference frame. $\sigma_{\l mn,2}$ includes the effects of the centrifugal force and ellipticity of the planet's figure as a result of rotation. In the limit of slow rotation, it is appropriate to truncate the expansion at first order in $\lambda$, in which case Equation~\ref{eq.correction_expansion} reduces to the well-known correction of \cite{1951ApJ...114..373L} in which the Coriolis force breaks the frequency's degeneracy with respect to the azimuthal order $m$.

Expressions for $\sigma_{\l mn,1}$ and $\sigma_{\l mn,2}$ are obtained through the perturbation theory; in practical terms they are inner products involving the zeroth-order eigenfunctions and operators describing the Coriolis and centrifugal forces and ellipticity. Corrections related to the distortion of equipotential surfaces require knowledge of the planetary figure as a function of depth, and these are provided directly by the theory of figures as described in \S\ref{s.gravity}.

This formalism is constructed to retain the separability of eigenmodes in terms of the spherical harmonics $Y_\l^m$, so that each corrected planet mode may still be uniquely specified by the integers $\l$, $m$ and $n$ and the expressions \ref{eq.eigenfunctions} and \ref{eq.displacement_eigenfunctions} hold for the corrected eigenfunctions.
Generally speaking, distinct modes whose frequencies are brought into close proximity by the perturbations from rotation may interact, yielding modes of mixed character. In the second-order theory applied to rigid rotation, selection rules limit these interactions to pairs of modes with the same $m$ and with $\l$ differing by $-2$, $0$, or $+2$. \cite{1981SvA....25..627V} found that for $f$- and $p$-modes with $\l\leq8$ these additional frequency perturbations do not exceed $0.5\%$, roughly an order of magnitude smaller than the second-order corrections themselves, and indeed generally smaller than the truncation error associated with neglecting higher-order correction terms (see below). There is thus little to be gained from incorporating mode-mode interactions given the accuracy of the present theory, but mode-mode interactions could be meaningfully taken into account in a third-order perturbation theory. The present work neglects mode-mode interactions.

Further details on the calculation of these rotation corrections are given by \cite{1990PhDT.........3M}, which the present implementation follows closely\footnote{\cite{1990PhDT.........3M} corrected several typographical errors from \cite{1981SvA....25..627V} and \cite{1981SvA....25..724V}, and one error was introduced: Equation (A1.27) for the ellipticity correction $I_5$ is missing a factor of two in the second term.}.
The interior density and sound speed discontinuities described above necessitate additional second-order corrections accounting for the ellipticity of these transitions and the gravitational potential perturbation felt throughout the planet as a result \citep[Section 5]{1981SvA....25..627V}.

Equation~\ref{eq.correction_expansion} provides the mode frequency as seen in inertial space.
This frequency can in turn be related to a pattern speed---the rotation rate of the full $m$-fold azimuthally periodic pattern---according to
    \begin{equation}
        \label{eq.pattern_speed}
        \ompat = \frac1m\siglmn,
    \end{equation}
which is suitable for direct comparison with the pattern speeds observed for waves in the rings.
For completeness, the mode frequency in the planet's corotating frame is related to the frequency seen in inertial space by
\begin{equation}
  \label{eq.corotating_frequency}
  \siglmn = \siglmnc + m\omegasat
\end{equation}
i.e., modes that are prograde in the planet's frame ($m>0$) modes appear to have larger frequencies in inertial space as a result of Saturn's rotation.

\begin{figure}[t]
	\begin{center}
        \includegraphics[width=\linewidth]{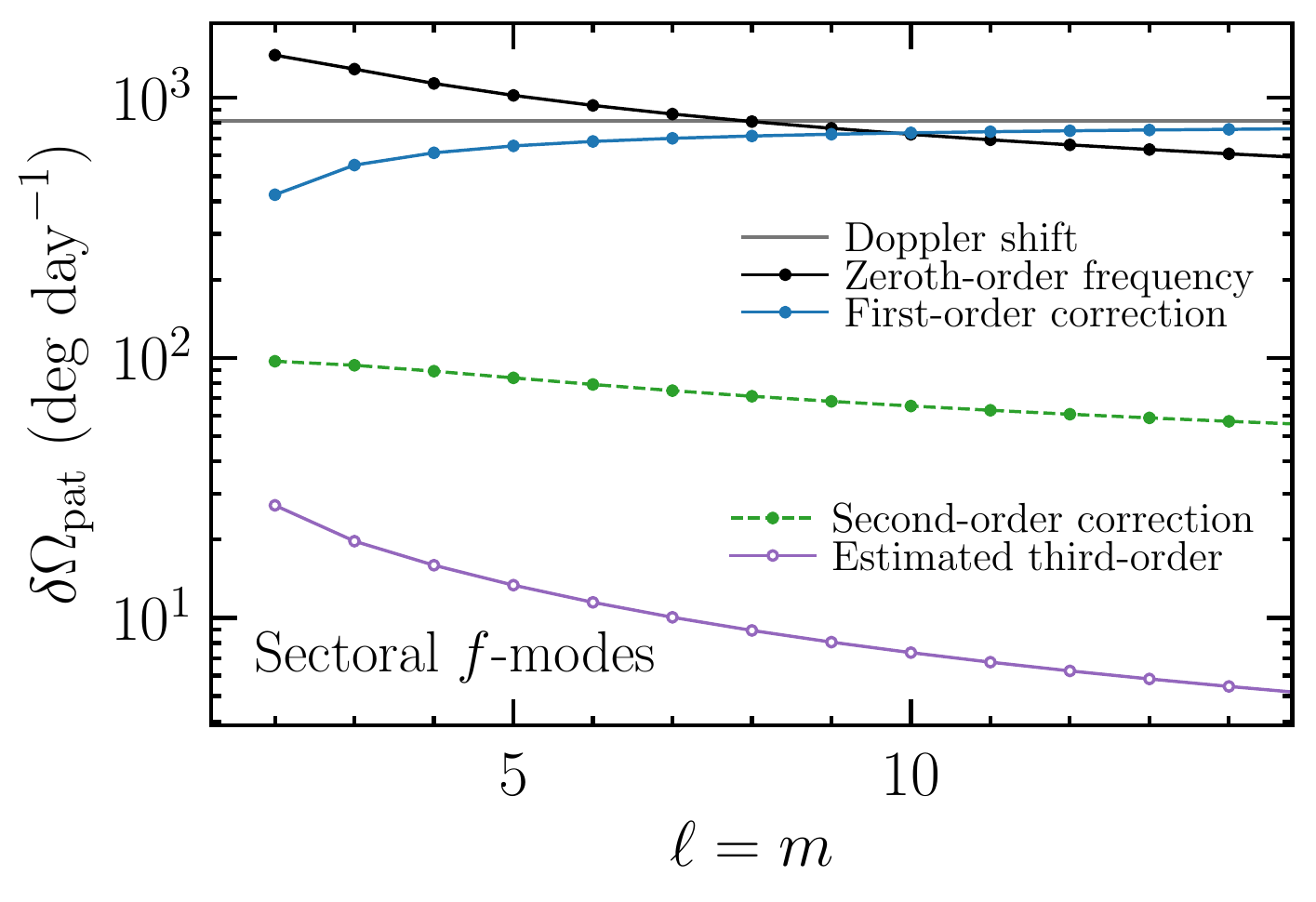}
		\caption{
        Magnitude of the contributions made to the modeled pattern speed by each of the four terms in Equation~\ref{eq.pattern_speed_contributions}, as well as the estimate (\ref{eq.third_order_term}) for the magnitude of third-order corrections. For these prograde modes the first-order corrections (Doppler plus Coriolis; blue solid curve) take positive values, the second-order corrections (centrifugal force and ellipticity; green dashed curve) take negative values, and the estimated third-order intrinsic corrections (purple solid curve) have no assumed sign.
        }
		\label{fig.pattern_speed_contributions}
	\end{center}
\end{figure}

As an illustration of the relative importance of these various contributions to the modeled pattern speed, we may substitute the frequency expansion~\ref{eq.correction_expansion} into \ref{eq.pattern_speed} to write
\begin{equation}
    \label{eq.pattern_speed_contributions}
    \ompat=\frac{\tildesigma}{m}+\frac{\sigma_{\l mn,1}\omegasat}{m}+\frac{\sigma_{\l mn,2}\omegasat^2}{m\tildesigma}.
\end{equation}
These three contributions are shown in Figure~\ref{fig.pattern_speed_contributions}, which demonstrates that the second-order rotation corrections affect the pattern speeds at the level of $\gtrsim50\ {\rm deg\ day}^{-1}$ for modes with $\ell$ below $15$. These corrections are thus essential for comparison with the observed wave pattern speeds, whose uncertainties are no larger than approximately $0.1\ {\rm deg\ day}^{-1}$ (P.D. Nicholson, private communication).

Higher order terms in the series expansion are potentially also significant. A third-order theory would include in the expansion~\ref{eq.pattern_speed_contributions} a $\lambda^3$ term
\begin{equation}
  \label{eq.third_order_term}
  \Omega_{\rm pat}^{(3)}\equiv\frac{\sigma_{\ell mn,3}\Omega_{\rm S}^3}{m\tildesigma^2},
\end{equation}
where the nondimensional prefactor $\sigma_{\ell mn,3}$ involves significant mathematical complexity \citep{1998A&A...334..911S,2008ChJAA...8..285K}.  To establish an upper limit for the magnitude of third-order corrections, noting that $\left|\sigma_{\ell mn,2}\right|<\left|\sigma_{\ell mn,1}\right|$ for all modes we consider,
we suppose that similarly $\left|\sigma_{\ell mn,3}\right|\leq\left|\sigma_{\ell mn,2}\right|$ and thus adopt $\left|\sigma_{\ell mn,3}\right|=\left|\sigma_{\ell mn,2}\right|$ as an upper limit.
The resulting upper limits on third-order contributions to $f$-mode pattern speeds are indicated in
Figure~\ref{fig.pattern_speed_contributions}, which demonstrates that the truncation error associated with our second-order theory may be as large as $30\ {\rm deg\ day}^{-1}$ for $\l=m=2$, but decaying with increasing $m$. As discussed \S\ref{s.results} below, these error estimates are taken into account in our analysis to ensure that the systematic dependence of the truncation error on $m$ does not bias our estimate of Saturn's bulk rotation rate.

\section{Saturnian $f$-modes in the rings}
\label{s.resonances}
This section briefly summarizes the formalism \citep{1993Icar..106..508M} connecting Saturn's nonradial oscillations with orbital resonances in the rings.
\subsection{Resonance conditions}
\label{s.resonance_conditions}
    The condition for a Lindblad resonance is \citep{1979ApJ...233..857G}
    \begin{align}
        \label{eq.lindblad_condition}
        m(\Omega-\ompat) = \pm q\kappa
    \end{align}
with the upper sign corresponding an inner Lindblad resonance (ILR) and the lower sign corresponding to an outer Lindblad resonance (OLR), and with $q$ a positive integer. Taking the lower sign in Equation~\ref{eq.lindblad_condition} to consider an OLR, it physically represents the condition that the perturbing pattern overtakes an orbiting ring particle once every $m/q$ epicycles. This prograde forcing in phase with the ring particles' epicycles leads to a deposition of angular momentum that may launch a spiral density wave propagating toward the planet, assuming self-gravity is the relevant restoring force. At an ILR an orbiting particle instead overtakes the slower perturbing pattern once every $m/q$ epicycles, leading to a removal of angular momentum that may launch a spiral density wave that propagates away from the planet. Such waves are common in Saturn's rings at mean motion resonances with Saturnian satellites.

Vertical resonances satisfy an analogous condition, namely that the perturbing pattern speed relative to the ring orbital frequency is simply related to the characteristic vertical frequency $\mu$ in the rings:
\begin{equation}
    m(\Omega-\ompat)=\pm b\mu,
    \label{eq.vertical_condition}
\end{equation}
where $b$ is a positive integer and the vertical frequency $\mu(r)$ in the ring plane can be obtained from \citep{1983Icar...53..185S}
\begin{equation}
    \mu^2+\kappa^2=2\Omega^2.
\end{equation}
As with Lindblad resonances, there exist both inner and outer vertical resonances (IVRs and OVRs), depending on the sign of $\Omega-\ompat$. Self-gravity waves excited at vertical resonances generally propagate in the opposite sense from those excited at Lindblad resonances, so that bending waves at IVRs propagate toward the planet and those at ILRs propagate away. IVRs are common in the rings as a result of Saturnian satellites, namely those whose inclinations provide resonant vertical forcing.

In the above, the positive integer $q$ or $b$ is sometimes referred to as the `order' of the resonance. This work focuses on first-order ($q=1$ or $b=1$) resonances; higher-order resonances are possible \citep{2014Icar..234..194M} but the wave structures they produce may destructively interfere (P.D.~Nicholson, private communication) and these resonances do not appear to need to be invoked to explain the present data (see \S\ref{s.results} below).
Furthermore, in what follows we limit our attention to OLRs and OVRs because in practice, the prograde $f$-modes of modest angular degree have pattern speeds that exceed $\Omega(r)$ throughout the C ring.

The orbital and epicyclic frequencies $\Omega$ and $\kappa$ for orbits at low inclination and low eccentricity can generally be written as a multipole expansion in terms of the zonal gravitational harmonics $J_{2n}$, namely
    \begin{align}
        \Omega^2(r)= \frac{\gmsat}{r^3}
        \Bigg[
        1+\sum_{n=1}^\infty A_{2n}J_{2n}\left(\frac{a}{r}\right)^{2n}
        \Bigg] \label{eq.orbit_omega_expansion}
    \end{align}
and
    \begin{align}
        \kappa^2(r)= \frac{\gmsat}{r^3}
        \Bigg[
        1+\sum_{n=1}^\infty B_{2n}J_{2n}\left(\frac{a}{r}\right)^{2n}
        \Bigg], \label{eq.orbit_kappa_expansion}
    \end{align}
with the $J_{2n}$ values scaled to the appropriate reference equatorial radius $a$. The $A_{2n}$ and $B_{2n}$ are rational coefficients and are tabulated by \cite{1988JGR....9310209N}. We use the even harmonics of \cite{iess2018} through $J_{12}$ for the purposes of locating resonances in the ring plane, although the gravity field only affects radial locations of resonances and has no bearing on $f$-mode pattern speeds. We therefore use the latter for quantitative comparison between model $f$-modes and observed waves.

The above relations constitute a closed system allowing the comparison of planet mode frequencies to the frequencies of waves observed at resonances in the rings. In cases where we do compare resonance \emph{locations}, the resonant radius for a Lindblad or vertical resonance is obtained by numerically solving Equation~\ref{eq.lindblad_condition} or \ref{eq.vertical_condition}.

\subsection{Which modes for which resonances?}
Each planet mode can generate one of either density waves or bending waves. The type of wave that the $\l mn$ mode is capable of driving depends on its angular symmetry, and in particular the integer $\l-m=(0, 1, 2, 3, \ldots)$. Modes with even $\l-m$ are permanently symmetric with respect to the equator, and so are not capable of any vertical forcing. However, they are antisymmetric with respect to their azimuthal nodes, and so do contribute periodic azimuthal forcing on the rings. The reverse is true of modes with odd $\l-m$, whose perturbations are antisymmetric with respect to the equator and so do contribute periodic vertical forcing on ring particles. Meanwhile their latitude-average azimuthal symmetry as experienced at the equator prevents them from forcing ring particles prograde or retrograde.

In what follows, we restrict our attention to prograde $f$-modes, namely the normal modes with $m>0$ and $n=0$. Acoustic modes with overtones ($n>0$; $p$-modes) are not considered because that they contribute only weakly to the external potential perturbation due to self-cancellation in the volume integral of the Eulerian density perturbation; see Equation~\ref{eq.phip_lmn_final_expression}. We further limit our consideration to prograde modes because while $f$-modes that are retrograde in the frame rotating with the planet can in principle be boosted prograde by Saturn's rotation (see Equation~\ref{eq.corotating_frequency}), we find that the resulting low pattern speeds ($\lesssim500\ {\rm deg\ day}^{-1}$) would place any Lindblad or vertical resonances beyond the extent of even the A or B rings. Finally, azimuthally symmetric ($m=0$) modes do not lead to Lindblad or vertical resonances.

\section{Results for rigid rotation}
\label{s.results}
\begin{figure*}[t]
	\begin{center}
        \includegraphics[width=2.0\columnwidth]{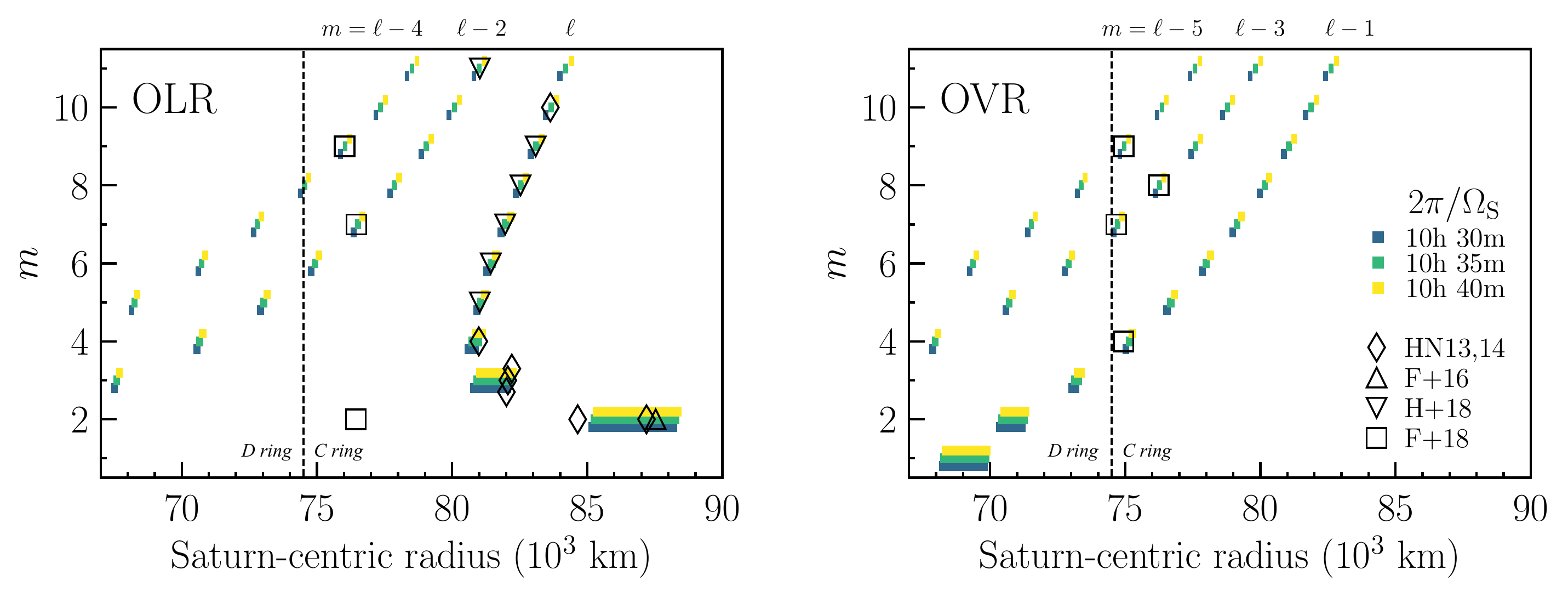}
		\caption{
        Locations of resonances with our model Saturn's $f$-modes (colorful horizontal spans) and wave features observed in Saturn's C ring using stellar occultations in \cassini VIMS data (open symbols; see references in Table~\ref{t.identifications}). The number of spiral arms $m$ (or equivalently, the azimuthal order of the perturbing planet mode) is shown versus distance from Saturn's center in the ring plane. \textit{Left panel:} Outer Lindblad resonances, which can excite inward-propagating spiral density waves in the rings. The three roughly vertical model sequences correspond to modes with $m=\l$, $m=\l-2$, and $m=\l-4$ from right to left. The three observed $m=3$ density waves are offset vertically for clarity. \textit{Right panel:} Outer vertical resonances, which can excite outward-propagating bending waves in the rings. The three vertical model sequences correspond to $m=\l-1$, $m=\l-3$, and $m=\l-5$ from right to left.
        Model resonances are colored by the assumed Saturn rotation rate as described in the legend; the resonances indicated for each rotation rate are slightly offset vertically for clarity.
        }
		\label{fig.locations}
	\end{center}
\end{figure*}

Figure~\ref{fig.locations} summarizes the OLR and OVR locations of prograde model Saturn $f$-modes with $\l-m$ between zero and five, together with locations of 17 inward-propagating density waves and four outward-propagating bending waves observed in \cassini VIMS data. A visual comparison in this diagram provides a strong indication that the $f$-modes are responsible for the majority of the wave features shown. In particular, we can make unambiguous identifications for the $f$-modes at the origin of 10 of the 17 density waves, and all four of the bending waves; these visual identifications are summarized in Table~\ref{t.identifications}.

The remaining seven density waves at $m=2$ and $m=3$ exhibit frequency splitting that is likely attributable to mixing with deep $g$-modes as proposed by \cite{2014Icar..242..283F}, and which our model, lacking a stable stratification outside the core, does not attempt to address. We thus omit all $m=2$ and $m=3$ waves from the quantitative analysis that follows, although we note that the predicted $\l=m=2$ and $\l=m=3$ $f$-mode OLR locations do generally coincide with the locus of observed density waves for these $m$ values, the sole exception being the close-in W76.44. This wave was only recently detected in VIMS data \citep{rgf2018}, and while coupling with deep $g$-modes is a possible interpretation (E.~Dederick, private communication), this wave may be particularly challenging to explain due to its large splitting from the other three $m=2$ waves.
We also note that the frequency and $m$ value of the outermost $m=2$ density wave in the ringlet within the Maxwell gap \citep{2016Icar..279...62F} were predicted by \cite{2014Icar..242..283F}.

As discussed in \S\ref{s.eigenstuff}, the density and sound speed discontinuities inherent to the three-layer interior structures assumed for Saturn affect the $f$-mode frequencies. Their effect is strongest for the lowest-degree $f$-modes, which have significant amplitude at these deep transitions. This is evident in Figure~\ref{fig.locations} in the considerable spread of predicted locations for resonances with the $\l=\{2,3\}$ $f$-modes. By $\l\gtrsim4$ the $f$-modes have low enough amplitudes at these deep density transitions that their frequencies are not strongly affected.

The model $f$-modes whose resonance locations coincide with the remainder of the observed waves contain a striking range of radial and latitudinal structures, including the rest of the sectoral $(\l=m)$ sequence up to $\l=m=10$, as well as seven non-sectoral ($\l\neq m$) modes with $\l-m=\{1,2,3,4,5\}.$ These waves are evidently the result of time-dependent tesseral harmonics resulting from Saturn's nonradial oscillations.

Although general agreement for these is evident at the broad scale of Figure~\ref{fig.locations}, the observed wave pattern speeds are known to a precision better than $0.1\ {\rm deg\ day}^{-1}$ for the weakest waves yet measured (P. D. Nicholson, private communication). This high precision warrants a closer inspection of the pattern speed residuals with respect to our predictions. What follows in the remainder of this section is an analysis of these residuals and their dependence on the assumed interior model and rotation rate.

\begin{deluxetable*}{clcllcllcc}
    \newcommand{\tup}{\bigtriangleup} 
    \newcommand{\tdown}{\bigtriangledown} 
    \newcommand{\dia}{\diamondsuit} 

\tabletypesize{\footnotesize}
\tablecolumns{10}
\tablewidth{1.0\columnwidth}
\tablecaption{\label{t.identifications}C ring wave patterns and Saturn $f$-mode associations}
\tablehead{
&&&\multicolumn{3}{c}{Observed}&&\multicolumn{3}{c}{Model prediction}\\
\cline{4-6}\cline{8-10}
\colhead{Reference\tablenotemark{a}} &
\colhead{Wave} &
\colhead{Symbol\tablenotemark{b}} &
\colhead{$m$} &
\colhead{$\ompat$\tablenotemark{c}} & 
\colhead{Type} &
&
\colhead{$\l$} &
\colhead{$\ompat$\tablenotemark{c}} &
\colhead{$\Delta\ompat\ ({\rm model-obs})$\tablenotemark{c}}
}
\startdata
F+18          &  W76.44\tablenotemark{\dag}   & $\square$ &  2    &   2169.3                          &   OLR &&   ---       &   ---                 &   --- \\
HN13          &  W84.64\tablenotemark{\dag}   & $\dia$    &  2    &   1860.8                          &   OLR &&   ---       &   ---                 &   --- \\
HN13          &  W87.19\tablenotemark{\dag}   & $\dia$    &  2    &   1779.5                          &   OLR &&   ---       &   ---                 &   --- \\
F+16          &  Maxwell\tablenotemark{\dag}  & $\tup$    &  2    &   1769.2                          &   OLR &&   ---         &   --- &   --- \\
HN13          &  W82.00\tablenotemark{\dag}   & $\dia$    &  3    &   1736.7                          &   OLR &&   ---       &   ---                 &   --- \\
HN13          &  W82.06\tablenotemark{\dag}   & $\dia$    &  3    &   1735.0                          &   OLR &&   ---       &   ---                 &   --- \\
HN13          &  W82.21\tablenotemark{\dag}   & $\dia$    &  3    &   1730.3                          &   OLR &&   ---         &   --- & --- \\
HN13          &  W80.98                       & $\dia$    &  4    &   1660.4                          &   OLR &&   4         &   $1657.87\textup{--}1673.41$ & $-2.49\textup{--} 13.05$ \\
H+18          &  W81.02a                      & $\tdown$  &  5    &   1593.6                          &   OLR &&   5         &   $1592.08\textup{--}1596.05$ & $-1.54\textup{--} 2.43$ \\
H+18          &  W81.43                       & $\tdown$  &  6    &   1538.2                          &   OLR &&   6         &   $1537.10\textup{--}1539.51$ & $-1.13\textup{--} 1.28$ \\
H+18          &  W81.96                       & $\tdown$  &  7    &   1492.5                          &   OLR &&   7         &   $1491.73\textup{--}1493.72$ & $-0.73\textup{--} 1.26$ \\
F+18          &  W76.46                       & $\square$ &  7    &   1657.7                          &   OLR &&   9         &   $1655.86\textup{--}1657.35$ & $-1.86\textup{--} -0.37$ \\
H+18          &  W82.53                       & $\tdown$  &  8    &   1454.2                          &   OLR &&   8         &   $1453.93\textup{--}1455.23$ & $-0.30\textup{--} 1.00$ \\
H+18          &  W83.09                       & $\tdown$  &  9    &   1421.8                          &   OLR &&   9         &   $1421.83\textup{--}1422.55$ & $-0.01\textup{--} 0.71$ \\
F+18          &  W76.02                       & $\square$ &  9    &   1626.5                          &   OLR &&   13        &   $1626.48\textup{--}1627.46$ & $-0.02\textup{--} 0.96$ \\
HN14          &  W83.63                       & $\dia$    &  10   &   1394.1                          &   OLR &&   10        &   $1394.03\textup{--}1394.71$ & $-0.03\textup{--} 0.65$ \\
H+18          &  W81.02b                      & $\tdown$  &  11   &   1450.5                          &   OLR &&   13        &   $1451.53\textup{--}1453.07$ & $1.04\textup{--} 2.58$ \\
\vspace{-3mm}\\
F+18          &  W74.93                       & $\square$ &  4    &   1879.6                          &   OVR &&   5         &   $1871.22\textup{--}1875.42$ & $-8.42\textup{--} -4.22$\\ 
F+18          &  W74.67                       & $\square$ &  7    &   1725.8                          &   OVR &&   10        &   $1723.99\textup{--}1725.28$ & $-1.77\textup{--} -0.48$ \\
F+18          &  W76.24                       & $\square$ &  8    &   1645.4                          &   OVR &&   11        &   $1644.89\textup{--}1645.81$ & $-0.54\textup{--} 0.38$ \\
F+18          &  W74.94                       & $\square$ &  9    &   1667.7                          &   OVR &&   14        &   $1667.72\textup{--}1668.85$ & $-0.01\textup{--} 1.12$
\enddata
\tablenotetext{a}{HN13 denotes \hnone, HN14: \hntwo, F+16: \french, H+18: \cite{2018arXiv181104796H}, F+18: \cite{rgf2018}.}
\tablenotetext{b}{cf. Figure~\ref{fig.locations}}
\tablenotetext{c}{${\rm deg\ day}^{-1}$}
\tablenotetext{\dag}{Member of a multiplet of waves of the same type having the same $m$ but different frequencies, possibly the result of resonant coupling between the $f$-mode of the same $m$ identified here and a deep $g$-mode as demonstrated by \cite{2014Icar..242..283F}. Thus no unambiguous identification with our pure $f$-mode predictions is possible. See discussion in \S\ref{s.results}; for the relevant $\l=m=2$ $f$-mode we predict $1743.34-1845.28\ {\rm deg\ day}^{-1}$ and for the $\l=m=3$ $f$-mode we predict $1729.29-1777.28\ {\rm deg\ day}^{-1}$.}
\end{deluxetable*}

\subsection{Saturn's seismological rotation rate}
\label{s.seismological_rotation}
Saturn's bulk rotation rate has to date been deduced from a combination of gravity field and radiometry data from the Pioneer, \voyager and \cassini spacecrafts (e.g., \citealt{1981GeoRL...8..253D}, \citealt{2005Sci...307.1255G}, \citealt{2006Natur.441...62G}, \citealt{2007Sci...317.1384A}).
Along different lines, \cite{2015Natur.520..202H} optimized interior models to the observed gravity field and oblateness to extract the rotation rate. Since we have demonstrated that the frequencies of Saturnian $f$-modes depend strongly on $\omegasat$ through the influence of the Coriolis and centrifugal forces and the ellipticity of level surfaces, a natural question is, what interior rotation rate is favored by the waves detected so far that appear to be associated with modes in Saturn's interior?

\begin{figure}[t]
	\begin{center}
        \includegraphics[width=\columnwidth]{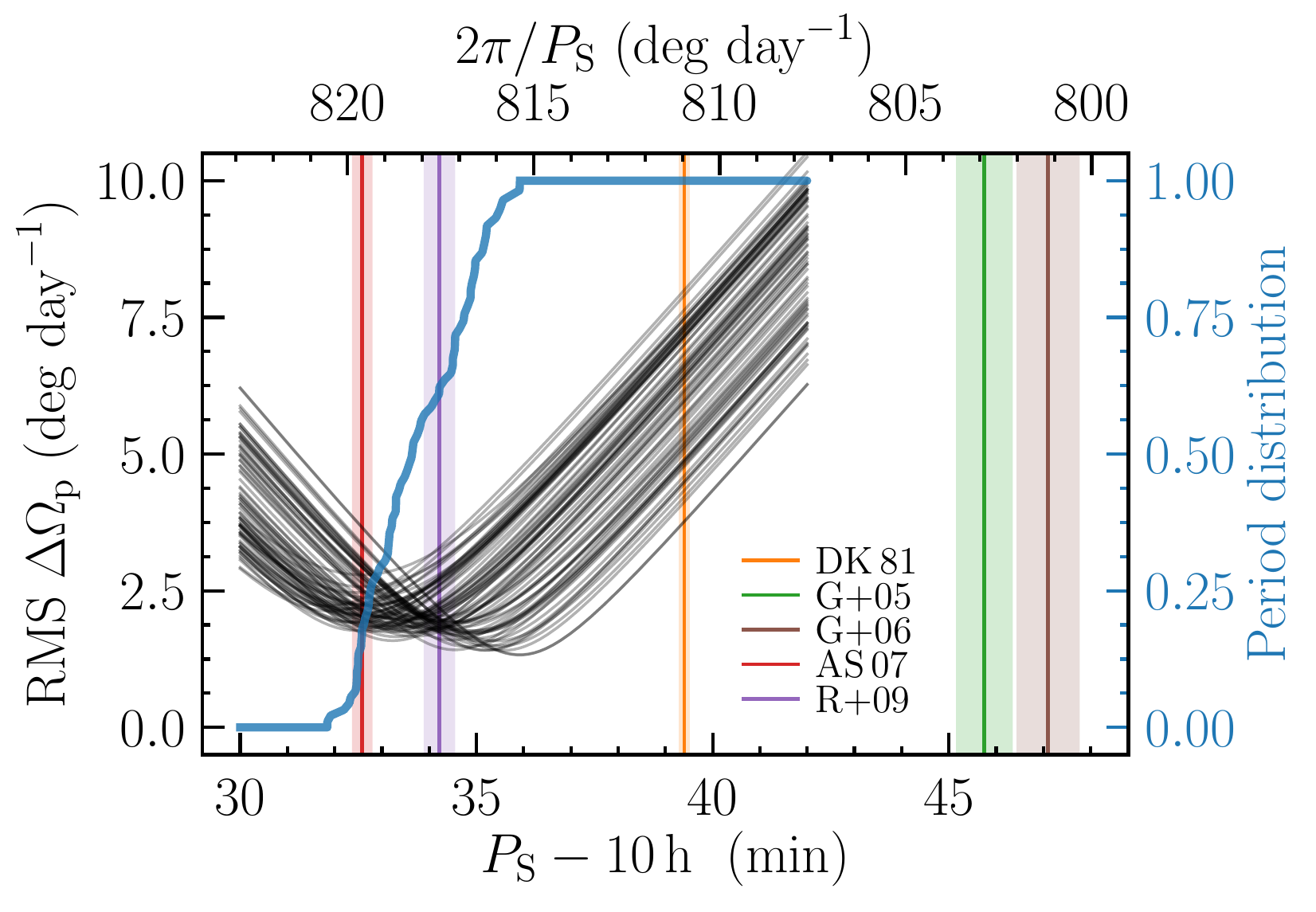}
		\caption{
        Saturn's rotation rate from fits to the set of observed C ring waves identified with Saturnian $f$-modes.
        RMS pattern speed residuals across the full set of waves are shown as a function of Saturn's assumed rotation rate. Each black curve corresponds to a single interior model from the sample shown in Figure~\ref{fig.profiles_compare_interiors}. The thick blue curve shows the cumulative distribution of rotation periods minimizing the weighted RMS pattern speed residuals for each model; its median and 5\%/95\% quantiles are given in \S\ref{s.seismological_rotation}.
        Vertical lines with shaded errors indicate Saturn rotation rates in the literature, references to which are given in the text. For visual clarity the \cite{2015Natur.520..202H} result of $10{\rm h}\,32{\rm m}\,(45\pm46){\rm s}$ referred to in the text is omitted from the diagram.
        }
		\label{fig.rotation}
	\end{center}
\end{figure}

Given an observed C ring wave with a pattern speed $\ompat^{\rm obs}$ that appears to be associated with a predicted Saturn model $f$-mode resonance with pattern speed $\ompat$ and azimuthal order matching the observed number of spiral arms $m$, we calculate the pattern speed residual $\Delta\ompat\equiv\ompat-\ompat^{\rm obs}$. For each Saturn interior model and rotation rate considered, we calculate a weighted root-mean-square (RMS) value of $\Delta\ompat$ over the set of mode-wave pairs according to
\begin{equation}
  \label{eq.weighted_rms}
  {\rm RMS}\ \Delta\ompat\equiv\left[\sum_i w_i\left|\Delta\Omega_{\rm pat,\, i}\right|^2\right],
\end{equation}
where the weights $w_i$ are assigned in inverse proportion with the maximum magnitude of third-order corrections as described in \S\ref{s.rotation_corrections}, the weights sum to unity, and $i$ indexes the set of waves that we have identified with Saturn $f$-modes, namely those with $\ell$ values and model pattern speeds listed in Table~\ref{t.identifications}. The resulting curves are shown in Figure~\ref{fig.rotation} for rotation periods between $10{\rm h}\,30{\rm m}$ and $10{\rm h}\,42{\rm m}$. The relation between ${\rm RMS}\ \Delta\ompat$ and $\omegasat$ always exhibits a distinct minimum, owing to the strongly correlated response of the $f$-mode frequencies to varying $\omegasat$. In particular, the predicted pattern speeds increase uniformly with faster Saturn rotation.

The optimal Saturn rotation period depends on the interior model chosen, as does the quality of that best fit: interior models favoring longer rotation periods generally achieve a slightly better bit. To account for this in our esimate of Saturn's bulk rotation period, we weight the optimized rotation period from each interior model in inverse proportion to the value of ${\rm RMS}\ \Delta\ompat$ obtained there. The cumulative distribution of rotation rates resulting from our sample of interior models is shown in Figure~\ref{fig.rotation}. This distribution may be summarized as
$\psat=10.561^{+0.031}_{-0.022}\,{\rm h}=10{\rm h}\, 33{\rm m}\, 38{\rm s}^{+1{\rm m}\, 52{\rm s}}_{-1{\rm m}\, 19{\rm s}}$ where the leading value corresponds to the median and the upper (lower) error corresponds to the 95\% (5\%) quantile. This may be expressed in terms of a pattern speed as $2\pi/\psat=818.13^{+2.41}_{-1.70}\,{\rm deg\ day}^{-1}$.

Although these seismological calculations vary the assumed rotation rate, the underlying interiors randomly sampled against $J_2$ and $J_4$ using the theory of figures as described in \S\ref{s.gravity} assumed the \citet{1981GeoRL...8..253D} \voyager rate, in principle an inconsistency of the model. As a diagnostic we generate a new sample from the gravity field, but adopting $\mrot=0.14201$ consistent with the 10.561h median rotation period derived here. Repeating the remainder of this analysis we find a very similar distribution of optimal rotation periods, the median shifting to longer periods by approximately one minute as a result of the slightly different interior mass distributions obtained. The frequencies of the $f$-modes themselves are inherently more sensitive to Saturn's assumed rotation rate than are the low-order gravity harmonics $J_2$ and $J_4$, a consequence of the $f$-modes extending to relatively high $m$ where Saturn's rotation imparts a larger fractional change to the frequency (see Figure~\ref{fig.pattern_speed_contributions}).

\subsection{Is rigid rotation adequate?}
\label{s.residuals}
\begin{figure*}[t]
	\begin{center}
        \includegraphics[width=\linewidth]{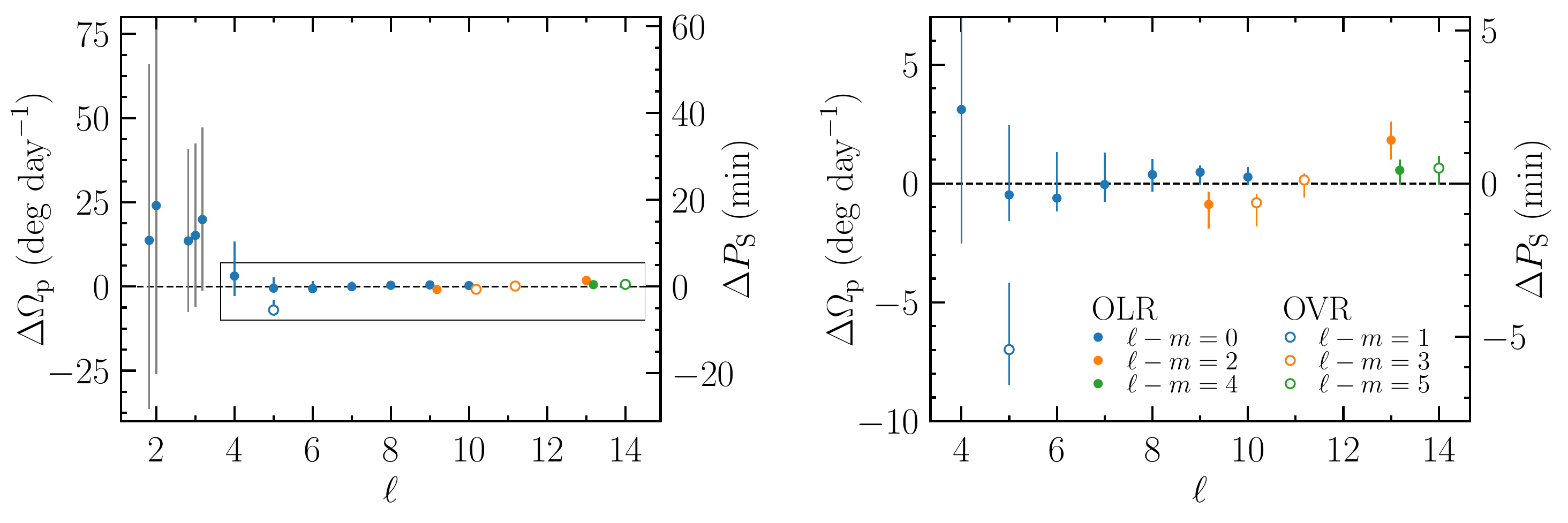}
		\caption{Pattern speeds residuals (predicted minus observed) for models each calculated at their optimal Saturn rotation period. \emph{Left panel:} All residual frequencies, including those for the $m=2$ and $m=3$ sectoral $f$-modes for which identification with specific $m=2$ or $m=3$ density waves is not possible. For these modes residuals are shown with respect to each of the nearby density waves having the correct $m$ value. The domain of the right panel is indicated. \emph{Right panel:} Frequency residuals for the 14 waves identified with Saturn $f$-modes and used to constrain Saturn's rotation. Circular markers are for one interior model randomly chosen from our sample, while vertical lines show the span of residuals obtained for the full sample. These vertical spans thus indicate the amount of freedom available from the low-order gravity field as applied to three-layer Saturn models, when the rotation rates are tuned using the seismology. Note that these spans do not represent random uncertainties because the residuals for the various modes are highly correlated. The vertical axis at right expresses the residuals in terms of minutes of Saturn rotation, i.e., the degree to which Saturn would need to be spun up or down to fit a given wave's observed pattern speed. Four pairs of modes that are members of same-$\l$ multiplets are evident (see discussion in \S\ref{s.discussion}); the pairs with $\l=9$, 10, and 13 are slightly offset horizontally for clarity.
        }
		\label{fig.residuals}
	\end{center}
\end{figure*}

The lack of any perfect fit among the range of interior structures and rotation rates we have considered is evident in Figure~\ref{fig.rotation}, where the RMS pattern speed residuals reach approximately 1.2 ${\rm deg\ day}^{-1}$ at best, an order of magnitude larger than the typical observational uncertainty of approximately $0.1\ {\rm deg\ day}^{-1}$ associated with even the weakest waves we compare to here (P.D. Nicholson, private communication). The absolute residuals are shown mode by mode in Figure~\ref{fig.residuals}, including the full span of residuals obtained over the sample of interior models, each one evaluated at its optimal rotation rate. Points lie on both sides of zero by construction, but again no model provides an entirely satisfactory fit.

First, it is notable that the model pattern speed covariance (the diagonal elements of which set the vertical spans in the residuals of Figure~\ref{fig.residuals}) varies so strongly and non-monotonically with $m$. This can be understood as a consequence of the tradeoff between the decreasing zeroth-order frequency and the increasing contribution from the first-order rotation correction with increasing $\l$, as can be seen from Figure~\ref{fig.pattern_speed_contributions} for the sectoral modes. At high $\l$, the zeroth-order frequency loses out to the first-order correction. Since the latter is proportional to $\omegasat$, the overall pattern speeds vary more strongly with rotation than at intermediate $\l$.
At low $\l$, where the frequency is dominated by the zeroth order contribution and so rotation plays a smaller role, the large model covariance is due mostly to sensitivity to the locations of the core boundary and envelope transition, sensitivity that decays rapidly with increasing $\l$ as modes are confined increasingly close to the planet's surface.

A significant observational development has been made by \cite{2018arXiv181104796H} in their detection of density waves corresponding to the full set of Saturn's sectoral $f$-modes from $\l=m=2$ up to $\l=m=10$, constituting frequency measurements for modes that possess the same latitudinal symmetry but sample an uninterrupted sequence of depths within Saturn.
On the other hand, the non-sectoral ($\l\neq m$) $f$-mode waves reported by \cite{2018arXiv181104796H} and \cite{rgf2018} extend the detections up to $\l=14$ but also importantly sample a variety of \emph{latitudinal} structures inside the planet by virtue of the range in their values of $\l-m$.
Thus in principle the available modes serve to constrain differential rotation inside Saturn.

With this in mind, the second panel of Figure~\ref{fig.residuals} is a valuable illustration because any strong differential rotation as a function of depth or latitude would generally manifest as systematic trends in the residuals $\Delta\ompat$ as a function of $\l$ or $\l-m$ respectively when referred to the rigid model. Instead, the residuals exhibit no obvious systematic dependence on $\l$, although small systematic departures as a function of $\l-m$ may indicate the presence of differential rotation as a function of latitude. In particular, in each of the four cases where two modes belonging to the same multiplet have been observed ($\l=5,\ 9,\ 10,\ {\rm and}\ 13$), the two frequencies are offset by between 1 and 5 deg day$^{-1}$.

More firm conclusions regarding the presence or strength of differential rotation are not possible given the present theoretical accuracy limitations discussed in \S\ref{s.rotation_corrections}. A more accurate treatment of rotation effects could potentially increase the predicted pattern speeds of the low-$m$ modes by as many as tens of $\rm deg\ day^{-1}$ (see Figure~\ref{fig.pattern_speed_contributions}) which could produce a spectrum consistent with a spin frequency increasing by several percent toward the planet's surface.
Indeed, this systematic uncertainty motivates the weighted fit that we carry out in our estimate of Saturn's bulk rotation in \S\ref{s.seismological_rotation}. Ultimately a more accurate perturbation theory, or else non-perturbative methods \citep[e.g.,][]{2018MNRAS.tmpL.212M}, will be required to fully interpret the implications for differential rotation inside Saturn.

\section{Strength of forcing}
\label{s.torques}
The adiabatic eigenfrequency calculation that forms the basis for this work provides no information about excitation or damping of normal modes, processes which have yet to be adequately understood in the context of gas giants.

Stochastic excitation of modes by turbulent convection such as in solar-type oscillations is one obvious candidate for Jupiter and Saturn, where convective flux dominates the intrinsic flux in each planet. However, the expectation from simple models for resonant coupling of $f-$ and $p-$modes with a turbulent cascade of convective eddies (e.g., \citet{2018Icar..306..200M} following the theory of \citet{1997IAUS..181..287K}) is that these modes are not excited to the amplitudes necessary to provide the mHz power excess that \cite{2011A&A...531A.104G} attributed to Jovian $p$-modes.

Recent work from \citet{2017ApJ...837..148D} demonstrated that a radiative opacity mechanism is not able to drive the Jovian oscillations, although they noted that driving by intense stellar irradiation is possible for hot Jupiters. \cite{2018ApJ...856...50D} and \cite{2018Icar..306..200M} each focused on water storms as a mode excitation mechanism, finding this too insufficient for generating a power spectrum akin to that reported by \cite{2011A&A...531A.104G}. \cite{2018Icar..306..200M} further demonstrated that deeper, more energetic storms associated with the condensation of silicates were viable.

In lieu of a complete understanding of the amplitudes of acoustic modes in gas giants, we simply adopt equal mode energy across the f-mode spectrum following
\begin{equation}
	\label{eq.amplitudes}
	E_{\l mn}\propto\sigma_{\l mn}^2|\bm \xi|^2=\rm constant,
\end{equation}
corresponding to the ``strong coupling'' case cited by \cite{1993Icar..106..508M}. Less efficient coupling of the turbulence with the $f$-modes could result in a steeper decline of equilibrium mode energy with frequency; \citet{1993Icar..106..508M} adopted $E_{\l mn}\propto\sigma_{\l mn}^{-13/2}$ as a limiting case.

Because the scaling relation \ref{eq.amplitudes} is only a proportionality, it remains to set an overall normalization by choosing the amplitude of a single mode. \cite{1993Icar..106..508M} proposed that the $\l=m=2$ $f$-mode OLR is the origin of the Maxwell gap, and accordingly anchored their amplitude spectrum by assuming that this mode had an amplitude sufficient to produce the OLR torque $T^L/\Sigma\sim10^{16}\ {\rm cm}^4\ {\rm s}^{-2}$ necessary to open a gap \citep{1991Icar...93....3R}. The corresponding displacement amplitude was of order 100 cm; we follow suit and adopt 100 cm as the amplitude of this
mode\footnote{While the connection between the $\l=m=2$ $f$-mode and the Maxwell gap itself has yet to be fully understood, it is tantalizing as this mode yields the largest gravity perturbations out of any of Saturn $f$-modes for any simple amplitude spectrum (see \S\ref{s.torques_subsection}). Furthermore, the ringlet within the gap harbors an $m=2$ density wave \citep{2016Icar..279...62F} as predicted from the Saturn mode spectrum calculated by \cite{2014Icar..242..283F}.}.

In what follows we normalize our $f$-mode eigenfunctions in accordance with the amplitude estimate of Equation~\ref{eq.amplitudes} and derive the resulting torques applied at OLRs and OVRs. While this amplitude law is but one of many plausible scenarios, any similar scaling relation will yield the same general dependence of Lindblad and vertical torques on $\l$, $m$, and position in the ring plane. In particular, the magnitudes of the torques decline monotonically with $\l$ for a given $\l-m$, and also with $\l-m$ for a given $m$. This is sufficient for a basic prediction of the \emph{relative} strengths of waves at the $f$-mode resonances calculated here, which will allow us to identify locations that may harbor hitherto-undetected waves.

\subsection{Torques and detectability}
\label{s.torques_subsection}
In deriving the magnitudes of torques applied at ring resonances we follow the approach of \citet{1993Icar..106..508M}. In a ring of surface mass density $\Sigma$, the linear torque applied at a Lindblad resonance is \citep{1979ApJ...233..857G}
\begin{equation}
    \label{eq.lindblad_torque}
    T_{\l m n}^L = -\frac{m\pi^2\Sigma}{\mathscr D_L}(2m+\l+1)\left(\Phi_{\l m n}^\prime\right)^2,
\end{equation}
where
\begin{equation}
    \label{eq.lindblad_d}
    \begin{split}
        \mathscr D_L = &-\left(3-\frac92J_2\left(\frac{a}{r_L}\right)^2\right)\Omega^2(1\mp m) \\
            & +\frac{21}{2}J_2\left(\frac{a}{r_L}\right)^2\Omega^2 + \mathcal O(J_2^2, J_4)
    \end{split}
\end{equation}
and $\Phi_{\l mn}^\prime$ is the magnitude of the perturbation to the gravitational potential caused by the $\l mn$ mode, evaluated at the Lindblad resonance $r=r_L$ in the ring plane $\cos\theta=0$.
Similarly the linear torque applied at a vertical resonance $r=r_V$ is \citep{1983Icar...53..185S,1993Icar..106..508M}
\begin{equation}
    \label{eq.vertical_torque}
    T_{\l mn}^V = \frac{m\pi^2\Sigma}{\mathscr D_V}\left(\frac{d\Phi_{\l mn}^\prime}{d\theta}\right)^2,
\end{equation}
where
\begin{equation}
    \label{eq.vertical_d}
    \begin{split}
        \mathscr D_V = &-\left(3+\frac92J_2\left(\frac{a}{r_V}\right)^2\right) \Omega^2(1\mp m) \\
            &-\frac{21}{2}J_2\left(\frac{a}{r_V}\right)^2\Omega^2 + \mathcal O(J_2^2, J_4)
    \end{split}
\end{equation}
and $({d\Phi_{\l mn}^\prime}/{d\theta})$ is to be evaluated at the vertical resonance $r=r_V$ and $\cos\theta=0$. In the expressions for $\mathscr D_L$ and $\mathscr D_V$ the upper (lower) signs correspond to inner (outer) Lindblad or vertical resonances, as in Equations~\ref{eq.lindblad_condition} and \ref{eq.vertical_condition}. An expression for $\Phi_{\l mn}^\prime$ is derived as in \citet{1993Icar..106..508M}; this is reproduced in Appendix \ref{s.potential_perturbations} for completeness. These expressions rely on integrals of the Eulerian density perturbation $\rho_{\l mn}^\prime$ over the volume of the planet. While accuracy to second order in Saturn's smallness parameter $\mrot$ would demand that this density eigenfunction include second-order corrections from the perturbation theory described in \S\ref{s.rotation_corrections}, the fact that only an order of magnitude calculation of the torques is required for the present purpose leads us to simply calculate these using the zeroth-order density eigenfunctions.

\begin{figure*}[t]
	\begin{center}
        \includegraphics[width=2.0\columnwidth]{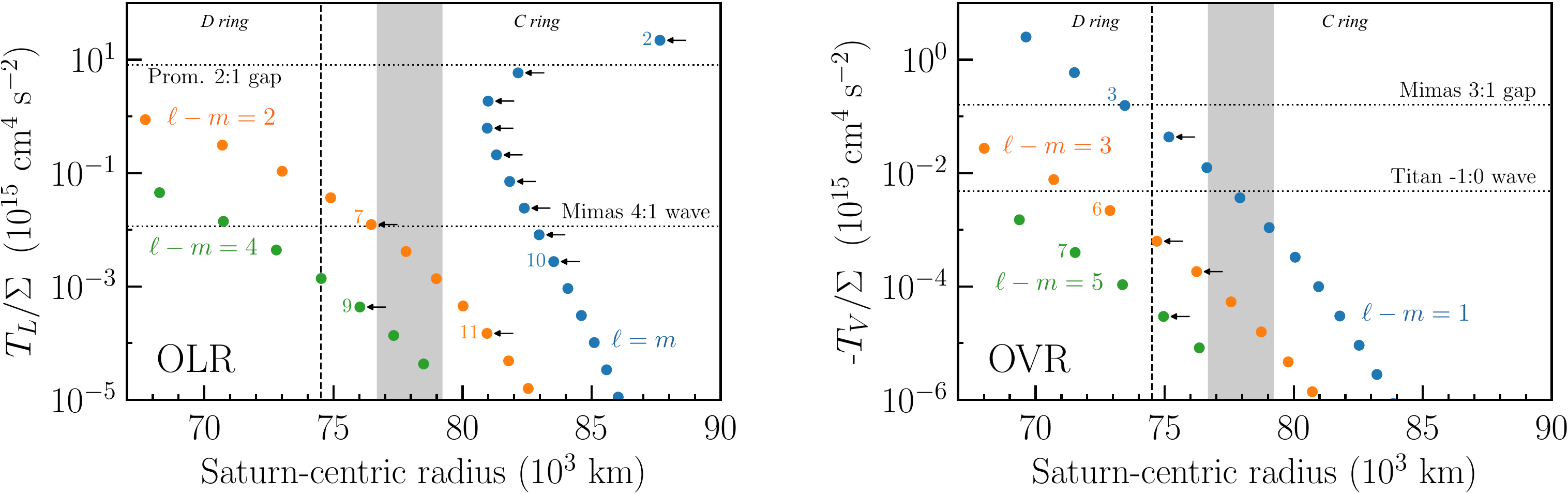}
		\caption{
        Strengths of torques per surface mass density exerted on the C and D rings by model Saturnian $f$-modes, with amplitudes assigned according to equal energy per mode following Equation~\ref{eq.amplitudes}. Model points (filled circles; shown for one randomly drawn interior model) are colored by their value of $\l-m$. Arrows highlight model $f$-modes that we have identified with observed waves as in Table~\ref{t.identifications}. The grey shaded region in both panels represents the region where maximum radial variations in ring orbits caused by the Titan 1:0 apsidal ILR exceed 300 m, making the detection of wave features more difficult; see \S\ref{s.torques}. A subset of resonances have been labeled to their left by their azimuthal wavenumber $m$ for ease of identification. Torques associated with waves or gaps at example satellite ILRs or IVRs in the C ring are indicated with dotted horizontal lines.
        }
		\label{fig.torques}
	\end{center}
\end{figure*}

To illustrate which modes are likely to excite the strongest ring features, Figure~\ref{fig.torques} summarizes the torques applied by the $f$-modes at OLRs and OVRs in the C and D rings assuming mode amplitudes follow equipartition per Equation~\ref{eq.amplitudes}. Because the torques (Equations~\ref{eq.lindblad_torque} and \ref{eq.vertical_torque}) are proportional to ring surface mass density $\Sigma$, itself strongly variable across the rings at a variety of spatial scales, we instead plot the normalized torques $T_L/\Sigma$ and $T_V/\Sigma$. These are straightforward quantities to calculate even with imperfect knowledge of the mass density itself. When comparing to detected wave patterns should be kept in mind that $\Sigma$ can play an important role in whether a given wave is likely to be driven to detectable amplitudes.

Saturnian waves can also be obscured by more prominent eccentric features, such as those associated with satellite resonances. Of particular importance is the strong Titan 1:0 apsidal resonance, which \citet{2014Icar..241..373N} studied in \cassini radio and stellar occultations and found responsible for driving the $m=1$ wave in the Titan/Colombo ringlet (77,879 km) and also dozens of other $m=1$ features from 74,000-80,000 km. Their test-particle model (cf. their Figure 19) predicts maximum radial deviations in excess of 100 m as much as 3,500 km away from that resonance, posing a serious challenge for the reconstruction of weaker wave features from stellar occultation profiles obtained at different phases. This substantial region of the C ring thus may be concealing waves driven at Saturn resonances, and Figure~\ref{fig.torques} accordingly indicates the region where the maximum radial deviations are larger than 300 m according to the model of \citet{2014Icar..241..373N}.

For context, the torques associated with four satellite resonances that open gaps or launch waves in the C ring are also shown in Figure~\ref{fig.torques}. Prometheus 2:1 ILR opens a gap in the C ring while the Mimas 4:1 ILR launches a wave. The Mimas 3:1 IVR opens a gap, while the Titan -1:0 nodal resonance launches a wave. Estimates for the strengths of these satellite torques are taken from \citet{1991Icar...93....3R,1991Icar...93...25R} and \cite{1993Icar..106..508M}.

\subsubsection{Conspicuously missing waves?}
\label{s.missing_waves}
Inspection of Figure~\ref{fig.torques} reveals a few $f$-mode resonances that this simple excitation model predicts to experience strong forcing, but where no waves have yet been detected. Four of the OLRs with $m=\l-2$ have normalized torques predicted to be greater than that of the detected $(\l,m)=(13,11)$ OLR. The most obvious of these is the $(8,6)$ OLR, which this model predicts to lie at 74,940 km, happening to be almost exactly coincident with the detected W74.93 and W74.94 OVR features \citep{rgf2018}. The fact that these OVR waves apparently dominate the signal at this position betrays some tension with the spectrum of amplitudes we have assumed, which predicts W74.93 and W74.94 to have torques one to three orders of magnitude lower than that predicted for the $(8, 6)$ OLR. Given the close proximity of these resonances, an appeal to the spatial dependence of $\Sigma$ seems unlikely to resolve this tension.

Of the remaining $m=\l-2$ OLRs stronger than $(13,11)$, none among $(10, 8)$, $(11, 9)$, or $(12,10)$ have had associated wave detections. This may be attributable to strong perturbations from the Titan apsidal resonance as discussed above. Among the predicted $m=\l-4$ resonances, the $(12, 8)$ OLR at 74,556 km is quite close to the inner boundary of the C ring where there are a series of gaps that have yet to be fully understood. Falling in such a gap could render such a resonance unobservable, although within the model uncertainty, this resonance could lie between the gaps or on gap edges.

As for the OVRs, the only resonances that yield waves that have been detected so far in the C ring are the four that fall closest to Saturn, and indeed the strongest predicted waves in each $\l-m$ have been observed. It warrants closer attention that three of the four strongest OVRs predicted in the C ring have not been associated with any wave feature, while waves have been observed at what should be weaker OVRs with $\l-m=3$ and $\l-m=5$. These three ``missing waves'' correspond to the $(\l, m)=(6,5)$, $(7,6)$, and $(8, 7)$ Saturn $f$-modes. Because of their location, it is possible that these waves are present but obscured by the Titan apsidal resonance.

To aid in the search for Saturnian resonances in the C ring, Table~\ref{t.missing_waves} lists the pattern speeds of all model OLRs and OVRs in the C ring with predicted torques comparable to or larger than than the smallest predicted torque associated with a wave that has already been observed. Likewise, Table~\ref{t.d_ring_waves} reports resonances predicted to lie in the D ring, although it is not clear whether any wave patterns there will ultimately be detectable given the ring's faintness.

\begin{deluxetable}{lllcc}
\tabletypesize{\footnotesize}
\tablecolumns{5}
\tablewidth{1.0\columnwidth}
\tablecaption{\label{t.missing_waves}Predicted OLRs and OVRs in the C ring without associated wave detections}
\tablehead{
\colhead{$\l$} &
\colhead{$m$} &
\colhead{Type} &
\colhead{$\ompat\ (\rm deg\ day^{-1})$} & 
\colhead{Remark (see \S\ref{s.torques})}
}
\startdata
11 &	    11 &	        OLR &	 $1368.5\textup{--}1371.5$ & 	\\
12 &	    12 &	        OLR &	 $1346.9\textup{--}1349.7$ & 	\\
13 &	    13 &	        OLR &	 $1327.7\textup{--}1330.1$ & 	\\
 8 &	     6 &	        OLR &	 $1742.1\textup{--}1747.6$ & 	Coincident with W74.93, W74.94\\
10 &	     8 &	        OLR &	 $1586.9\textup{--}1591.0$ & 	Near Titan apsidal\\
11 &	     9 &	        OLR &	 $1532.9\textup{--}1536.4$ & 	Near Titan apsidal\\
12 &	    10 &	        OLR &	 $1488.4\textup{--}1491.6$ & 	\\
14 &	    12 &	        OLR &	 $1419.3\textup{--}1421.8$ & 	\\
12 &	     8 &	        OLR &	 $1695.6\textup{--}1699.5$ & 	Among gaps\\
14 &	    10 &	        OLR &	 $1568.6\textup{--}1571.6$ & 	Near Titan apsidal\\
15 &	    11 &	        OLR &	 $1521.4\textup{--}1524.1$ & 	Near Titan apsidal\\
\vspace{-3mm} \\
6 &	     5 &	        OVR &	   $1737.0\textup{--}1743.6$ & Near Titan apsidal\\
7 &	     6 &	        OVR &	   $1646.4\textup{--}1652.1$ & Near Titan apsidal\\
8 &	     7 &	        OVR &	   $1578.0\textup{--}1582.8$ & \\
9 &	     8 &	        OVR &	   $1523.9\textup{--}1528.1$ & \\
10 &	     9 &	        OVR &	 $1479.8\textup{--}1483.5$ & \\
11 &	    10 &	        OVR &	 $1443.0\textup{--}1446.3$ & \\
12 &	     9 &	        OVR &	 $1581.1\textup{--}1584.6$ & Near Titan apsidal\\
13 &	    10 &	        OVR &	 $1530.1\textup{--}1533.1$ & Near Titan apsidal\\
15 &	    10 &	        OVR &	 $1604.7\textup{--}1607.6$ &
\enddata
\tablecomments{Pattern speeds can be mapped to physical locations given Saturn's equatorial radius and $J_{2n}$ using the relations in \S\ref{s.resonance_conditions}.}
\end{deluxetable}

\begin{deluxetable}{lllc}
\tabletypesize{\footnotesize}
\tablecolumns{4}
\tablewidth{1.0\columnwidth}
\tablecaption{\label{t.d_ring_waves}Predicted OLRs and OVRs in the D ring}
\tablehead{
\colhead{$\l$} &
\colhead{$m$} &
\colhead{Type} &
\colhead{$\ompat\ (\rm deg\ day^{-1})$}
}
\startdata
5 &	     3 &	        OLR &	   $2314.3\textup{--}2324.3$ 	 \\
6 &	     4 &	        OLR &	   $2034.3\textup{--}2042.2$ 	 \\
7 &	     5 &	        OLR &	   $1861.1\textup{--}1867.6$ 	 \\
9 &	     5 &	        OLR &	   $2063.1\textup{--}2069.4$ 	 \\
10 &	     6 &	        OLR &	 $1901.5\textup{--}1906.8$ 	 \\
11 &	     7 &	        OLR &	 $1784.6\textup{--}1789.0$ 	 \\
\vspace{-3mm} \\
2 &	     1 &	        OVR &	   $3359.2\textup{--}3400.7$ 	 \\
3 &	     2 &	        OVR &	   $2423.7\textup{--}2433.4$ 	 \\
4 &	     3 &	        OVR &	   $2061.8\textup{--}2070.8$ 	 \\
7 &	     4 &	        OVR &	   $2177.3\textup{--}2185.2$ 	 \\
8 &	     5 &	        OVR &	   $1968.1\textup{--}1974.5$ 	 \\
9 &	     6 &	        OVR &	   $1826.1\textup{--}1831.4$ 	 \\
11 &	     6 &	        OVR &	 $1970.5\textup{--}1975.6$ 	 \\
12 &	     7 &	        OVR &	 $1841.5\textup{--}1845.9$ 	 \\
13 &	     8 &	        OVR &	 $1743.8\textup{--}1747.5$
\enddata
\end{deluxetable}

\section{Discussion}
\label{s.discussion}
This work offers interpretations for the set of inward-propagating density waves and outward-propagating bending waves observed in Saturn's C ring in terms of resonances with Saturnian $f$-modes. It also demonstrates that Saturn's rotation state is of critical importance for Saturn ring seismology, a fact made evident by the systematic mismatch with the observed pattern speeds of these waves obtained assuming that Saturn rotates rigidly at the \voyager System III magnetospheric period of \citet{1981GeoRL...8..253D} or slower (see Figure~\ref{fig.rotation}). The interior configurations considered to arrive at this conclusion accounted somewhat generously for the freedom in the low-order gravity field, because the likelihood function used to obtain our posterior distribution of interior models assumed an inflated variance on $J_2$ to accord with the numerical precision of our theory of figures implementation (see \S\ref{s.gravity}). Because the resulting distribution included a diversity of heavy element and helium distributions, envelope transition locations, and core masses, the seismology suggests a tension with the \voyager rotation rate commonly assumed for Saturn's interior that different three-layer interior models seem unlikely to resolve. This conclusion based on the ring seismology adds support to the notion that periodicities in Saturn's magnetospheric emission \citep[e.g.,][]{1981GeoRL...8..253D,2005Sci...307.1255G,2006Natur.441...62G} may not be consistently coupled to the rotation of Saturn's interior \citep[e.g.,][]{2007Sci...316..442G,2009Natur.460..608R}.

The present model is potentially oversimplified in two major ways. First, the model is not suited to address the close multiplets of waves observed to have the same azimuthal wavenumber $m$, namely the multiplets of waves in the C ring with $m=2$ and $m=3$. The bulk of these seem naturally explained by the model of \citet{2014Icar..242..283F}, wherein avoided crossings between the $f$-modes and deep $g$-modes of higher angular degree give rise to a number of strong perturbations with the same $m$ value. However, in the wealth of new OLR and OVR wave patterns that have been measured from increasingly low signal to noise VIMS data since \citet{2014MNRAS.444.1369H}, it seems that only two waves add to the mixed-mode picture, both with $m=2$: the close-in W76.44 wave, and the Maxwell ringlet wave whose frequency and $m$ number were predicted by \cite{2014Icar..242..283F}.
The $f$-modes of higher angular degree have less amplitude in the deep interior and so are less likely to undergo degenerate mixing with any deep $g$-modes strongly. Indeed, there is not yet any direct evidence for $f$-modes with $\l>3$ undergoing avoided crossings with deep $g$-modes, although the outlying $(5,4)$ OVR warrants closer scrutiny in the mixed-mode context.

The second major simplification of the present model is the assumption that Saturn rotates rigidly. While upper limits can be established for the depth of shear in Jupiter or Saturn's envelopes on magnetohydrodynamic grounds \citep{2008Icar..196..653L,2017Icar..296...59C}, evidence gathered from spacecraft indicate that zonal wind patterns do penetrate to significant depths \citep{1982Sci...215..504S,2018Natur.555..223K}.
It has been proposed that the insulating molecular regions of these planets may be rotating differentially on concentric cylinders \citep{1976Icar...29..255B,1982Icar...52...62I,1986Icar...65..370I}, the zonal winds being the surface manifestation of these cylinders of constant angular velocity.
The mode identifications made in \S\ref{s.results} and Table~\ref{t.identifications} reveal that the seismological dataset now samples a variety of radial (via the angular degree $\l$) and latitudinal (via the latitudinal wavenumber $\l-m$) structures within Saturn and so should strongly constrain differential rotation in Saturn's interior. If our rigid model systematically underpredicted $f$-mode frequencies toward high $\l$, this would indicate that Saturn's outer envelope rotates faster than the bulk rotation. Such a result would be qualitatively consistent with the expectation for rotation on cylinders or an eastward equatorial jet that extends to significant depth, as well as with the rotation profiles that \cite{iess2018} deduced from the \cassini Grand Finale gravity orbits.
As discussed in \S\ref{s.residuals}, the lack of any such obvious systematic dependence of wave pattern speed residuals on $\l$ (see Figure~\ref{fig.residuals}) offers a preliminary indication that Saturn does \emph{not} experience strong differential rotation as a function of radius within the volume sampled by the $\l\geq4$ $f$-modes considered in this analysis, although we emphasize that the inclusion of higher order rotation corrections is necessary to confirm this.

The modes identified here also contain four instances of a pair of modes belonging to the same multiplet, i.e., a pair described by the same angular degree $\ell$ but different azimuthal order $m$. This carries significance for the prospect of deducing Saturn's rotation profile from the frequency splitting within each multiplet, although the important centrifugal forces and ellipticity due to Saturn's rapid rotation complicates the picture compared to the first-order rotation kernels commonly applied to helioseismology \citep{2003ARA&A..41..599T} and asteroseismology \citep[e.g.,][]{2012Natur.481...55B}.
 The frequency offsets that remain between modes with the same $\l$ but different $\l-m$ may point to a latitude-dependent spin frequency, although the manner in which this would fit in with a radius-independent spin frequency is unclear. Quantitative constraints on differential rotation via the $f$-modes awaits future work.


\section{Conclusions}
\label{s.conclusion}
We have presented new Saturn interior models and used them to predict the frequency spectrum of Saturn's nonradial acoustic oscillations. Comparison with waves observed in Saturn's C ring through \cassini VIMS stellar occultations reveals that the majority of these waves that are driven at frequencies higher than the ring mean motion are driven by Saturn's fundamental acoustic modes of low to intermediate angular degree $\l$.

The frequencies of Saturn's $f$-modes probe not only its interior mass distribution, but also its rotation state, especially those modes of higher $\l$. We used the frequencies of the observed wave patterns to make a seismological estimate of Saturn's rotation period assuming that it rotates rigidly. Using these optimized models, we proposed that small but significant residual signal in the frequencies of the observed waves as a function of $\l$ suggests that Saturn's outer envelope may rotate differentially, although we are unable to draw quantitative conclusions given the accuracy with which the present theory accounts for rotation in predicting the $f$-mode frequencies.

Saturn ring seismology is an interesting complement to global helioseismology, ground-based Jovian seismology, and asteroseismology of solar-type oscillators. Because the rings are coupled to the oscillations purely by gravity, they are fundamentally sensitive to the modes without nodes in the density perturbation as a function of radius, and the observation of modes from $\l=2$ to $\l\sim15$ stands in contrast with helioseismology where the vast majority of detected modes are acoustic overtones ($p$-modes) and $f$-modes only emerge for $\l\gtrsim100$ \citep[e.g.,][]{2008JPhCS.118a2083L}. Likewise ground-based Jovian seismology accesses the mHz-range $p$-modes and Saturn ring seismology fills in the picture for frequencies down to $\sim100\ \mu{\rm Hz}$. Because of their point-source nature, main sequence and red giant stars with \emph{CoRoT} and \emph{Kepler} asteroseismology means that typically only dipole ($\l=1$) or quadrupole ($\l=2$) modes are observable because of geometric cancellation for modes of higher $\l$ \citep{2013ARA&A..51..353C}. In contrast, the proximity of the C and D rings to Saturn renders them generally sensitive also to higher $\l$ so long as the modes exhibit the correct asymmetries. We finally reiterate that Saturn is a rapid rotator ($\omegasat/\sigma_0\sim0.4$), more in line with pulsating stars on the upper main sequence \citep{1998A&A...334..911S} than with stars with \emph{CoRoT} and \emph{Kepler} asteroseismology, and to our knowledge this is the most complete set of modes characterized to date for such a rapidly rotating hydrostatic fluid object.

This work buttresses the decades-old hypothesis \citep{stevenson1982} that Saturn's ordered ring system acts as a sensitive seismograph for the planet's normal mode oscillations. The set of Saturnian waves detected in the C ring so far thus provide important contraints on Saturn's interior that are generally independent from those offered by the static gravity field. Future interior modeling of the solar system giants will benefit from joint retrieval on the gravity harmonics and normal mode eigenfrequencies.

\acknowledgements
We thank Philip Nicholson and Matthew Hedman for extensive discussions about the detection and characterization of waves in the occultation data, Nadine Nettelmann for invaluable guidance in the theory of figures, and the anonymous referee for thoughtful comments that greatly improved the quality of the paper. C.M. further thanks Andrew Ingersoll, Stephen Markham, Ethan Dederick, Jim Fuller, and Daniel Thorngren for helpful conversations. This work was supported by NASA through Earth and Space Science Fellowship program grant NNX15AQ62H to C.M. and \cassini Participating Scientist program grant NNX16AI43G to J.J.F.  The University of California supported this work through multi-campus research award 00013725 for the Center for Frontiers in High Energy Density Science. Some of these calculations made use of the Hyades supercomputer at UCSC, supported by NSF grant AST-1229745 and graciously mantained by Brant Robertson.

\appendix

\section{Perturbations to the external potential}
\label{s.potential_perturbations}
The density perturbations associated with nonradial planet oscillations generally lead to gravitational perturbations felt outside the planet. These perturbations can be understood as time-dependent components to the usual zonal and tesseral gravity harmonics, and these are derived here following \citet{1993Icar..106..508M}.

As in the standard harmonic expansion for the static gravitational potential outside an oblate planet \citep{1978ppi..book.....Z}, the time-dependent part of the potential arising from nonradial planet oscillations can be expanded as
\begin{align}
    \label{eq.time_dependent_potential_expansion}
    \begin{split}
    \Phi^\prime(t) =
    \frac{GM}{r}\sum_{n=0}^\infty
    \Bigg\{
    -&\sum_{\l=2}^\infty\left(\frac{a}{r}\right)^\l J_{\l n}^\prime P_\l(\cos\theta)
    +\sum_{\l=2}^\infty\sum_{m=-\l}^\l\left(\frac{a}{r}\right)^\l P_\l^m(\cos\theta)
    \left[C_{\l mn}^\prime\cos m\phi+S_{\l mn}^\prime\sin m\phi\right]
    \Bigg\}.
    \end{split}
\end{align}
The coefficients $J_{\ell n}^\prime$, $C_{\ell n}^\prime$ and $S_{\ell n}^\prime$ are analogous to the usual gravity harmonics, but with the background density replaced by the Eulerian density perturbation $\rhop(\mathbf r, t)$ due to the oscillation in the $\l mn$ mode:
\begin{align}
    \begin{split}
        Ma^\l J_{\l n}^\prime &= -\int\rhop_{\l mn}(\mathbf r, t)\,r^\l P_\l(\cos\theta)\,d\tau \label{eq.coefficients_initial_expression}, \\
        Ma^\l \left(\begin{array}{cc}C_{\l mn}^\prime \\ S_{\l mn}^\prime\end{array}\right) &= \frac{2(\l-m)!}{(\l+m)!}\int\rhop_{\l mn}(\mathbf r, t)\,r^\l P_\l^m(\cos\theta)\left(\begin{array}{cc}\cos m\phi \\ \sin m\phi\end{array}\right)\,d\tau, \\
    \end{split}
\end{align}
where $d\tau=r^2\sin\theta\,d\theta\,d\phi\,dr$ is the volume element and the integrals are carried out over the volume of the planet. Given that our solutions for the density perturbation take the form
\begin{align}\begin{split}
    \rhop_{\l mn}(\mathbf r, t)&=Y_\l^m(\theta,\phi)\rhop_{\l n}(r)e^{-i\siglmn t} \\
    &= c_0 P_\l^m(\cos\theta)\rhop_{\l n}(r)e^{i(m\phi-\siglmn t)},
\end{split}\end{align}
where
\begin{align}
    c_0\equiv (-1)^{\frac{m+|m|}{2}}\left[\frac{2\l+1}{4\pi}\frac{(\l-|m|)!}{(\l+|m|)!}\right]^{1/2}, 
\end{align}
the integrals in Equation~\ref{eq.coefficients_initial_expression} are separable:
\begin{align}
    Ma^\l J_\l^\prime &= -c_0 e^{-i\sigma t}
        \int_0^{2\pi}e^{im\phi}\,d\phi
        \int_0^\pi \left[P_\l(\cos\theta)\right]^2\sin\theta\,d\theta
        \int_0^a\rhop_{n\l}(r)\,r^{\l+2}\,dr \label{eq.js_separated} \\
    Ma^\l \left(\begin{array}{cc}C_{\l m}^\prime \\ S_{\l m}^\prime\end{array}\right) &= \frac{2(\l-m)!}{(\l+m)!}c_0 e^{-i\sigma t}
        \int_0^{2\pi}e^{im\phi}\left(\begin{array}{cc}\cos m\phi\\\sin m\phi\end{array}\right)\,d\phi
        \int_0^\pi\left[P_\l^m(\cos\theta)\right]^2\,\sin\theta\,d\theta
        \int_0^a\rhop_{n\l}(r)\,r^{\l+2}\,dr.\label{eq.cs_ss_separated}
\end{align}
Notice from the symmetric integrand over azimuth that the $J_\l^\prime$ only have contributions from axisymmetric ($m=0$) modes, while the $C_{\l m}^\prime$ and $S_{\l m}^\prime$ only have contributions from nonaxisymmetric ($m\neq0$) modes. Using the orthogonality of the associated Legendre polynomials
\begin{align}
    \int_0^\pi P_\l^m(\cos\theta)P_{\l^\prime}^{m^\prime}(\cos\theta)\,\sin\theta\,d\theta =
    \int_{-1}^1 P_\l^m(\mu)P_{\l^\prime}^{m^\prime}(\mu)\,d\mu=
    \frac{2\delta_{\l\l^\prime}\delta_{mm^\prime}}{(2\l+1)}\frac{(\l+m)!}{(\l-m)!},
    \label{eq.legendre_theta_orthogonality}
\end{align}
Equations \ref{eq.js_separated} and \ref{eq.cs_ss_separated} reduce to
\begin{align}
    Ma^\l J_\l^\prime
        &= -\left(\frac{4\pi}{2\l+1}\right)^{1/2}e^{-i\siglmn t}
        \int_0^a\rhop_{\l n}(r)\,r^{\l+2}\,dr, \\ 
    Ma^\l C_{\l m}^\prime
        &= (-1)^{\frac{m+|m|}{2}}\left[\frac{4\pi}{(2\l+1)}\frac{(\l-|m|)!}{(\l+|m|)!}\right]^{1/2}
        e^{-i\siglmn t}
        \int_0^a\rhop_{\l n}(r)\,r^{\l+2}\,dr, \\ 
    S_{\l m}^\prime &= iC_{\l m}^\prime. 
    \label{eq.coefficients_final_expressions}
\end{align}
The coefficients $S_{\l m}$ are identical to the $C_{\l m}$ up to a phase offset and can thus be ignored. These expressions for the coefficients $J_\l^\prime$ and $C_{\l m}^\prime$ can be substituted into the expansion~\ref{eq.time_dependent_potential_expansion} to write the $\l m n$ component of the external potential perturbation as
\begin{equation}
    \label{eq.phip_lmn_final_expression}
    \Phi_{\l m n}^\prime(\mathbf r, t)=
    \left\{
    \begin{array}{ll}
        \displaystyle\frac{G}{r^{\l+1}}P_\l(\cos\theta)\left(\frac{4\pi}{2\l+1}\right)^{1/2}e^{-i\siglmn t}\int_0^a\rho_{\l m n}^\prime(r)r^{\l +2}\,dr, & m=0, \\
                \displaystyle\frac{G}{r^{\l+1}}P_\l^m(\cos\theta)(-1)^{\frac{m+|m|}{2}}\left[\frac{4\pi}{(2\l+1)}\frac{(\l-|m|)!}{(\l+|m|)!}\right]^{1/2}e^{-i\siglmn t}\cos m\phi\,\int_0^a\rho_{\l mn}^\prime(r)\,r^{\l+2}\,dr, & m\neq0.
    \end{array}
    \right.
\end{equation}
As above, we restrict our attention to prograde f-modes, namely those normal modes having $m>0$ and $n=0$. Thus for the modes of interest the amplitude of the potential perturbation felt at a radius $r$ outside Saturn is simply
\begin{equation}
    \label{eq.potential_perturbation_magnitude_prograde_fmodes}
    \Phi_{\l m0}^\prime(r,\theta)=\frac{G}{r^{\l+1}}P_\l^m(\cos\theta)(-1)^{\frac{m+|m|}{2}}\left[\frac{4\pi}{(2\l +1)}\frac{(\l-|m|)!}{(\l+|m|)!}\right]^{1/2}\int_0^a\rho_{\l m0}^\prime(r)\,r^{\l+2}\,dr
\end{equation}
where the time dependence and azimuthal dependence are omitted for the purposes of estimating the magnitudes of torques on the rings.

\software{\gyre \citep{2013MNRAS.435.3406T}, \emcee \citep{2013PSP..125..306F}, Matplotlib \citep{Hunter:2007}, SciPy \citep{scipy}, NumPy \citep{numpy}}
\facilities{ADS}


\newcommand{\noop}[1]{}

\end{document}